\documentclass[pdflatex,sn-nature]{sn-jnl}% Style for submissions to Nature Portfolio journals
\usepackage{lineno}
\usepackage{graphicx}%
\usepackage{multirow}%
\usepackage{amsmath,amssymb,amsfonts}%
\usepackage{amsthm}%
\usepackage{rsfso}%
\usepackage[title]{appendix}%
\usepackage{xcolor}%
\usepackage{textcomp}%
\usepackage{manyfoot}%
\usepackage{booktabs}%
\usepackage{listings}%

\theoremstyle{thmstyleone}%
\theoremstyle{thmstyletwo}%

\theoremstyle{thmstylethree}%
\newcommand{\edit}[1]{{#1}} % Makes the argument bold
\unnumbered

\begin{document}

\title[Quantum-Accurate Vibrational and Conformational Dynamics from Small Molecules to Proteins with Machine-Learned Force Fields]{\edit{Quantum-Accurate Vibrational and Conformational Dynamics from Small Molecules to Proteins with Machine-Learned Force Fields}}
\author[1]{\fnm{Sergio} \sur{Suárez-Dou}}\email{sergio.suarezdou@uni.lu}
\author[1]{\fnm{Miguel} \sur{Gallegos}}\email{miguel.gallegosgonzalez@uni.lu}
\author[1]{\fnm{Kyunghoon} \sur{Han}}\email{kyunghoon.han@uni.lu}
\author[1]{\fnm{Florian N.} \sur{Brünig}}\email{florian.bruenig@uni.lu}
\author[1]{\fnm{Joshua T.} \sur{Berryman}}\email{josh.berryman@uni.lu}
\author*[1]{\fnm{Alexandre} \sur{Tkatchenko}}\email{alexandre.tkatchenko@uni.lu}
\affil[1]{Department of Physics and Materials Science, University of Luxembourg, L-1511 Luxembourg City, Luxembourg}

\abstract{
\edit{
Biomolecular thermodynamics and spectroscopy depend on relative conformer energies, local curvatures, and collective dipole fluctuations on the potential-energy surface. Conventional molecular mechanics force fields enable large-scale simulations, but their fixed functional forms can misrepresent infrared intensities, mode character, and environment-dependent vibrational response. Here we assess general-purpose machine-learned force fields across small molecules, finite-temperature infrared spectra, gas-phase peptides, and monomeric, oligomeric, and solvated protein assemblies.  To enable this analysis, we introduce QVib, a dataset of 293 molecules and 1365 conformers, together with peptide amide-band benchmarks and p53 oligomerization-domain models, to evaluate vibrational transferability from DFT references to experimental spectra. Across these systems, machine-learned force fields substantially improve over molecular mechanics in reproducing DFT-level forces, vibrational frequencies, densities of states, mode eigenvectors, conformational energetics, and experimental infrared spectra. Among models with explicit long-range electrostatics, SO3LR provides the most favourable accuracy--cost balance for the biomolecular systems considered. These results show that machine-learned force-field dynamics can recover collective, environment-dependent vibrational landscapes at near-DFT fidelity, enabling spectroscopically validated biomolecular simulations at force-field-like cost.}
}

\maketitle

\section{Introduction}

Our understanding of biomolecular structural dynamics has been mainly obtained through static crystal structures, empirical force fields \cite{Hollingsworth2018, Ciccotti2022}, and, more recently, sequence-to-structure tools such as AlphaFold \cite{Jumper2021}. These approaches have provided invaluable insights into the architecture and local flexibility of proteins, often portraying them as relatively rigid entities with dynamics confined to localized fluctuations around a stable fold \cite{Chan-Yao-Chong2023}. Molecular dynamics (MD) simulations driven by empirical potentials have extended this view by permitting disordered or highly complex motion to be investigated, yet the prevailing paradigm still emphasizes localized motions and modular behavior, often by necessity in order to focus investigation on a region of interest, for example a drug binding pocket. A growing body of experimental evidence suggests that dynamic characteristics away from the better-characterised situations to which classical force fields are targeted may be more complex~\cite{Salvi2019, Marcellini2020, Salvi2022, Bolik-Coulon2022, Love2023,  Stroet2024, Nettels2024}.\\

Beyond localized fluctuations, biomolecular systems explore a hierarchy of metastable states separated by free-energy barriers, which govern slow conformational transitions and functional dynamics. Within each metastable basin, the potential-energy surface (PES) is often well approximated by the harmonic approximation, enabling normal mode analysis (NMA) to extract local curvatures (frequencies) and eigenvectors that encode the principal collective motions. Although this approximation neglects anharmonic couplings and large-amplitude motions, it links local curvature to thermodynamic quantities such as entropic contributions from the vibrational density of states and to spectroscopic observables, since infrared (IR) and Raman intensities depend on mode-resolved dipole and polarizability derivatives \cite{Ditler2022}. The validity of the harmonic approximation must be assessed to determine whether macroscopic observables, such as transition rates and temperature-dependent conformational changes, can be described accurately.\\

One critical limitation of conventional MD approaches based on classical force fields is their neglect of quantum-mechanical effects. Quantum phenomena such as polarization, charge transfer, and anharmonic couplings are not only essential for accurate energy landscapes but also manifest directly in vibrational spectra, where subtle shifts and intensity variations encode information about electronic structure and environmental response. Consequently, molecular mechanics force fields (MMFFs) can only be rigorously compared to experiment in regimes where such quantum contributions are negligible, a condition rarely satisfied in biologically relevant systems.\\

To understand biomolecular stability and vibrations from first principles, we compare widely used MMFFs with modern machine-learned force fields (MLFFs) trained exclusively on quantum-mechanical forces and energies. While MMFFs typically treat two-body electrostatics at all spatial distances using Ewald-like summations, they usually cover non-Coulomb two-body interactions only up to a distance of 8-10 {\AA} and many-body interactions only for the nearest two or three chemical bonding partners.  In contrast, modern MLFFs based on equivariant graph neural networks cover interactions up to $\approx$15 {\AA} as an `all-body' treatment.\\

State-of-the-art general-purpose MLFFs, such as \edit{MACE-off23 \cite{Kovacs2025MACEOFF}, MACE-POLAR-1 \cite{Batatia2026MACEPOLAR1}, ANI-2x \cite{Devereux2020ANI2x}, AIMNet2\cite{Anstine2025AIMNet2}, UMA \cite{Wood2025UMANeurIPS}, and SO3LR \cite{Kabylda2024}, use} a medium-range many-body description of the enviroment. \edit{Within them, SO3LR, AIMNet2, and MACE-POLAR-1 treat explicitly long-range electrostatic and dispersion terms. This design enables the model to account for interactions beyond short-range effects, improving accuracy for larger systems. Input training data for MLFF models is typically DFT forces, which can be improved with other observables, such as atomic volumes or molecular dipole moments from the same calculations \cite{Batatia2026MACEPOLAR1, Anstine2025AIMNet2, Kabylda2024}.} The larger ``effective neighbourhoods" of MLFFs permit the capture of more complex structure from the training data of a wide range of short and medium-range quantum effects such as polarization, charge transfer, and exchange repulsion, which are actively modulated by the local chemical environment. These effects are essential for the accurate modeling of reactive and electronically complex systems. Between the simplified picture of a pure classical MMFF and the immensely more complex parameter space of an MLFF, `polarizable' MMFFs attempt to model the lability of electron distributions, often in a first-order or linearised way, capturing some but not all of the important environment-dependent physics, especially as related to electrostatic moments \cite{Ren2003}. MLFFs mark a substantial improvement over conventional and also polarizable MMFFs by capturing non-local interactions and delicate local electronic effects with high accuracy \cite{Chmiela2018, Sauceda2019, Unke2021}.\\

It is important to recognize that interatomic interactions arising from electronic correlations can be even more farsighted than current MLFFs can describe. For example, atomic interactions in nanoscale systems can span distances of 10--100 nanometers \cite{Ambrosetti2016,Hauseux2020,Hauseux2022,Sosa2025} and even collective van der Waals interactions between small proteins and water have been shown to exhibit effective decays of 25-30 {\AA} \cite{Stohr2019}. We leave further investigation of such ultra-long-ranged effects on biomolecular dynamics to future work and here focus on analyzing complex quantum-mechanical effects at short and medium distances.\\

We begin by validating \edit{the aforementioned MLFF} on small molecules through NMA, comparing to the General Amber Force Field 2 (GAFF2) \cite{He2020}\edit{, broadly used in the MMFFs community for small molecule parametrization}.  \edit{Vibrational frequencies and mode-resolved dipole derivatives are benchmarked directly against the DFT references, with the PBE0+MBD \cite{Perdew1996, Adamo1999, Tkatchenko2012} and $\omega$B97M-V \cite{Mardirossian2016} exchange-correlation functionals used to train the models to ensure a fair comparison. IR spectra beyond the harmonic approximation are examined to assess the ability of the methods to capture anharmonic effects and intensity variations.} Building on this foundation, we present a detailed case study of $o$F‑Phe+H\textsuperscript{+}, where the computed PES of conformer isomers provides a mechanistic interpretation of IR spectral features. The scope is then expanded to peptide and protein systems, employing widely used empirical force fields—AMBER \cite{Lindorff-Larsen2010, Maier2015,Tian2020}, CHARMM36m \cite{Huang2017}, OPLS \cite{Robertson2015}, and AMOEBA \cite{Shi2013, Zhang2018}. \edit{For these larger molecules, SO3LR is, to our knowledge, the only currently available general-purpose MLFF efficient enough to make simulations at this scale practical while retaining the explicit treatment of long-range interactions and the level of fidelity required here.} Validation is extended to the Alanine-\edit{based peptides}, where NMA reveals quantum-level correspondence. \edit{Ala\textsubscript{5, 10, 15}LysH+ were used to directly compare IR prediction to SO3LR}. Finally, the p53 \edit{oligomerization} domain is investigated in monomeric and tetrameric forms, both in vacuum and solvated environments, to probe solvent and intraprotein effects on vibrational dynamics. These biologically critical and structurally complex systems are chosen to showcase the capture of environmental effects and dynamical couplings by MLFF that are inaccessible to more conventional approaches.\\

The comprehensive assessment of \edit{MLFFs and their} applicability across a wide variety of biochemical systems and configurations demonstrates its efficiency, accuracy, scalability, and transferability for the study of biomolecular stability and vibrations compared to high-fidelity quantum reference and experimental data when it is available. This paves the way towards fully predictive molecular dynamics simulations, including intrinsically disordered proteins and antibodies containing non-natural amino acids.

\section{Results and Discussion}

\subsection{Vibrational Landscapes of Small Molecules}
\begin{figure}
\includegraphics[width=\linewidth]{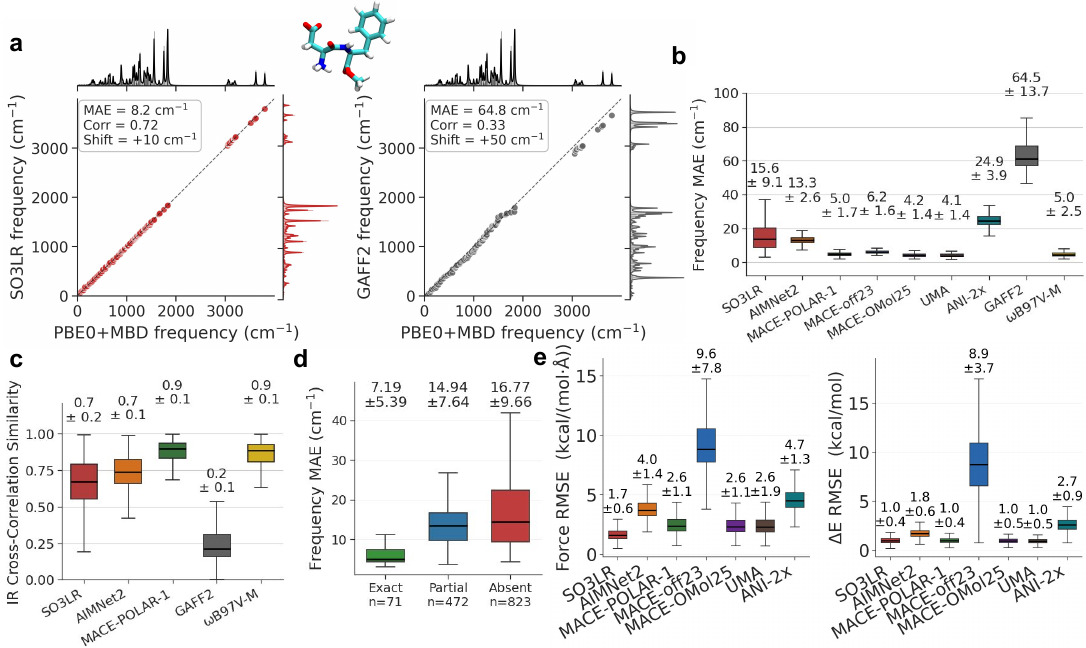}
\centering
\caption{\textbf{Vibrational benchmark of \edit{MLFFs} and GAFF2 using DFT as reference calculation.} \edit{\textbf{a} SO3LR (left) and GAFF2 (right) frequency correlation against PBE0+MBD for aspartame (Aspartame conformer 1 from QVib). The top and right marginals show Lorentzian-broadened IR spectra (FWHM = 15 cm\textsuperscript{-1}) of the reference and the method. Each panel reports the frequency MAE  (cm\textsuperscript{-1}) and the IR spectral cross-correlation similarity and best frequency shift  (cm\textsuperscript{-1}).
\textbf{b-e} QVib (n=1365 conformers from 293 molecules) whole set frequency MAE (\textbf{b}, cm\textsuperscript{-1}) and IR cross-correlation similarity (\textbf{c}, see \hyperref[cross_corr]{\S Cross correlation metrics}), comparison of MLFFs, GAFF2 and $\omega$B97M-V versus PBE0+MBD. 
\textbf{d} QVib SO3LR frequency MAE (cm\textsuperscript{-1}) stratified by the degree of overlap with the training set. Molecules are grouped into three categories: exact overlap (Exact, Tanimoto=1), partial overlap (Partial, Tanimoto$>$0), and no overlap (Absent, Tanimoto=0). \textbf{e} QM7-X per-molecule error distributions versus PBE0+MBD. Left: force RMSE (kcal\,mol\textsuperscript{-1}\,\AA\textsuperscript{-1}), right: $\Delta E$ RMSE (kcal\,mol\textsuperscript{-1}). Boxes span the interquartile range (Q1–Q3), the horizontal line marks the median, and whiskers extend to 1.5 × IQR; outliers are not shown. Numerical annotations above each box give the per-method mean ± standard deviation. $\omega$B97M-V is included as a DFT-level cross-functional reference. ANI-2x n=987 it is not been parametrize for P.}
Source data are provided as a Source Data File.}
\label{fig:small_mol}
\vspace{-4.0mm}
\end{figure}
%AT: Figure 1 is too packed. I suggest removing the second panel of c and second panel of d; and first panel of e (MAE and RMSE plots are very similar). You should also specify the dataset names on the plots c,d,e. Give a name to your 293 molecule dataset. [DONE]

As an initial probe of MLFFs as reporters of the molecular vibrational landscape, \edit{we assessed a representative set of general-purpose MLFFs, including MACE-off23 \cite{Kovacs2025MACEOFF}, MACE-POLAR-1 \cite{Batatia2026MACEPOLAR1}, ANI-2x \cite{Devereux2020ANI2x}, AIMNet2\cite{Anstine2025AIMNet2}, UMA \cite{Wood2025UMANeurIPS}, and SO3LR \cite{Kabylda2024}, and evaluated their ability to reproduce vibrational properties of small molecules against MMFFs and reference DFT calculations. Reference vibrational data were taken from PBE0+MBD \cite{Perdew1996, Adamo1999, Tkatchenko2012} and $\omega$B97M-V \cite{Mardirossian2016} calculations, providing a consistent quantum-mechanical baseline for the comparison}. In this section the MMFF used was GAFF2 \cite{He2020}\edit{, as an example of empirical force field for small molecules}. GAFF2 has transferable bond stretching, angle bending, and torsional parameters which have been developed evolutionarily over many years, based on a combination of empirical data and post-Hartree-Fock calculations, with semi-empirical AM1-BCC \cite{Jakalian2002} typically used to assign fractional atomic point charges as a per-molecule precalculation step \cite{Wang2004}. The \edit{QVib} benchmark set was designed as \edit{293 molecules with a total of 1365 conformers spanning common functional groups and 8 atomic elements (H, C, O, N, P, S, F, Cl). The set includes both relatively simple species, expected to lie closer to the training distributions of many general-purpose MLFFs, and more complex drug-like molecules such as ibuprofen, haloperidol, and aspartame, which provide a more stringent test of transferability. This composition allows us to assess model performance across regimes ranging from near-interpolative chemical environments to cases requiring greater extrapolation.}\\

IR spectroscopy is a widely used technique for probing molecular vibration \textit{via} its coupling to the electric field of light. In MD simulations, these vibrational properties emerge from the interatomic forces defined by the force field. Finite-temperature IR spectra can be computed from the dipole–dipole autocorrelation function obtained from MD trajectories, providing a test of the modelling of dynamics and electrostatics.
Unfortunately, the high computational cost of DFT-based MD makes it impractical to use DFT to take \textit{ab initio} IR spectra as a reference. As a cheaper alternative zero-temperature IR spectra can be approximated by harmonic vibrational analysis at optimized minimum-energy geometries (via diagonalisation of a mass-weighted Hessian), which is being successfully applied to model experimental spectra in materials and biomolecules with the PBE0+MBD exchange-correlation functional \cite{Hoja2019ReliablePolymorphs, Hoja2024, Boziki2024, Hunnisett2024, Rossi2014, Schubert2015}\edit{, which we adopt here as the reference level for IR calculations}. The lower cost of MLFFs such as SO3LR\edit{, AIMNet2 or MACE-POLAR-1} and of MMFFs makes it feasible to perform full MD simulations, allowing for a more realistic and dynamic representation of finite temperature vibrational behavior and dipole fluctuations (\hyperref[IR]{\S Infrared spectrum}). Below we investigate zero-temperature fluctuations \edit{(coupled to IR when possible)} as NMA (Fig.~\ref{fig:small_mol}).\\

\edit{As an illustrative example, Aspartame (Fig.~\ref{fig:small_mol}a \& see Sup. Fig. A1) highlights the qualitative and quantitative gap between MLFFs and a GAFF2. Harmonic analysis at the optimized minimum shows that SO3LR reproduces the reference vibrational frequencies with a mean absolute error (MAE) of 8.2~cm$^{-1}$, whereas GAFF2 exhibits substantially larger deviations (MAE = 64.8~cm$^{-1}$; Fig.~\ref{fig:small_mol}b). Importantly, the improved agreement is not limited to the vibrational frequencies. The cross correlation similarity (\hyperref[cross_corr]{\S Cross correlation metrics}) of the broadened IR spectra indicates that SO3LR also preserves the character of the IR-active normal modes, i.e., the dipole response associated with the vibrational displacements, whereas GAFF2 shows clear mismatches in the relative peak intensities and overall spectral pattern.}\\

\edit{Across the full benchmark, this behavior is systematic (Fig.~\ref{fig:small_mol}b-c): all tested MLFFs yield consistently lower frequency errors and closer agreement with the reference IR spectra than GAFF2. This indicates that MLFFs reproduce not only DFT-level curvature information, as reflected in the vibrational frequencies, but also the character of the IR-active normal modes through a more accurate dipole-response pattern. Within the MLFFs, the MACE family provides the best overall agreement with the reference across both frequency MAE and IR-spectrum similarity metrics. This strong performance should be interpreted in the context of training-set overlap: approximately \(\sim\)90\% (about \(\sim\)257 out of 293) of the molecules in this benchmark have close chemical analogues in the datasets used to train MACE-off23 (Sup.~Fig.~A2), placing most of the test set within an interpolative regime for this model, and the MACE family.}
\\
%AT: how many of the 293 molecules are in the datasets used to train MACE? This needs to be discussed right here. [DONE]

\edit{By comparison, SO3LR exhibits larger variability in performance across the dataset. This variability correlates with differences in training-set coverage: while MACE retains close training analogues for most molecules considered here, SO3LR lacks similar coverage for approximately \(\sim\)100 molecules (Sup.~Fig.~A3), a regime in which reduced predictive accuracy is observed. As shown in Fig.~\ref{fig:ala15}d, SO3LR yields robust vibrational predictions when molecular environments are fully or partially represented in the training data, whereas performance slightly degrades for molecules that lie farther from the training distribution.} \\

\edit{These results provide encouraging news for current general-purpose MLFFs. Across chemically diverse molecules, the different MLFFs show broadly consistent energy predictions and substantially improved force and vibrational accuracy compared with conventional MMFFs. The larger deviations observed for forces and vibrational frequencies are expected, since these quantities probe local derivatives of the potential-energy surface and are therefore more sensitive to the quality of the learned curvature than relative energies alone. Nevertheless, even in these more demanding metrics, the MLFFs remain markedly closer to the DFT reference than GAFF2. The present benchmark should therefore not be interpreted primarily as a ranking of model architectures, but as a diagnostic of the regimes in which current MLFFs already provide near-DFT fidelity and of the conditions where additional training coverage would be most beneficial. This is further illustrated by the QM7-X dataset \cite{Hoja2021}, which differs from equilibrium small-molecule datasets by containing a large number of systematically displaced, non-equilibrium geometries generated around equilibrium structures. QM7-X therefore probes the ability of the models to reproduce forces and energy variations away from the equilibrium manifold, where errors are expected to increase. Even in this more demanding regime, however, MLFF remains reliable to reference DFT calculations (Fig.~\ref{fig:small_mol}e). This is especially true for MLFFs that have been trained on large datasets (millions of molecules) of well converged DFT calculations.}
%AT: I think the above paragraph should be written in a much more optimistic way. It is clear that different models are very consistent in terms of energies. The forces and vib freq deviations are larger, but not as much as for MMFFs. This is really a very positive message for GP MLFFs. This should be highlighted and discussed more here. [DONE + QM7-X explanation]
\\

\begin{figure}
\centering
\includegraphics[width=\textwidth]{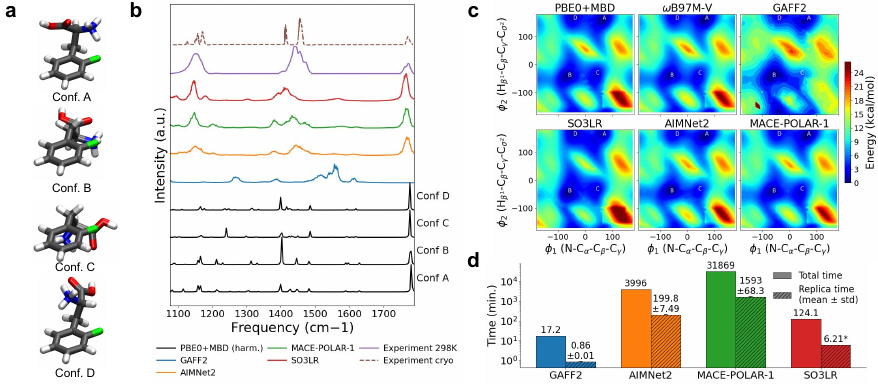}
\caption{\textbf{Vibrational analysis of L-$o$F-phenylalanine+H\textsuperscript{+}.} \textbf{a} Four stable conformational isomers, ordered from most to least stable, used as initial geometries for harmonic approximation and molecular dynamics simulations.\edit{ \textbf{b} Experimental cryo (dotted brown) and room temperature (purple) IR spectrum  \cite{Safferthal2023} compared with SO3LR (red), MACE-POLAR-1 (green),  AIMNet2 (orange), and GAFF2 (blue) spectra obtained from dipole autocorrelation MD, alongside harmonic approximation results from PBE0+MBD (black). \textbf{c} Potential energy surface mapped as a function of $\phi_1$ (N–C$_\alpha$–C$_\beta$–C$_\gamma$) and $\phi_2$ (H$_{\beta^1}$–C$_\beta$–C$_\gamma$–C$_{\sigma^2}$) dihedral angles. Letters indicate conformers defined in panel A. \textbf{d} Total GPU time required to obtain IR spectra from dipole‑autocorrelation MD (solid bars) and the corresponding per‑replica cost (striped bars), with standard deviations. *SO3LR simulations were executed in parallel using JAX parallelization; the replica time for SO3LR is the total time divided by 20 replicas. All timings correspond to runs on Nvidia Tesla V100 SXM2 GPUs provided by the Uni.lu HPC facility.} \edit{The IR} spectrum used a correction factor of 0.966, as in the reference. Source data are provided as a Source Data File.}
\label{fig:of-phe}
\vspace{-4.0mm}
\end{figure}

Trends in the finite-temperature IR are illustrated in L-$o$F-phenylalanine+H\textsuperscript{+} ($o$F-Phe+H\textsuperscript{+}).  $o$F-Phe+H\textsuperscript{+} exhibits a wide conformational diversity, driven by strong intramolecular interactions defining 4 stable isomers (Fig.~\ref{fig:of-phe}a). As shown by M.~Safferthal \textit{et al.}~\cite{Safferthal2023}, the middle-frequency region (1000–2000 cm\textsuperscript{-1}) strongly depends on the balance of these conformational isomers. We therefore investigate the four conformations separately by harmonic approximation at DFT reference level, then further probe the ensemble finite temperature spectrum comparing SO3LR, \edit{AIMNet2, MACE-POLAR-1} and GAFF2 versus experiment (Fig.~\ref{fig:of-phe}b). A closer examination of the PES obtained with different methods (Fig.~\ref{fig:of-phe}c and See. Sup. Fig. 4A) highlights that \edit{MLFFs} closely reproduces the PBE0+MBD \edit{and $\omega$B97M-V} reference data, whereas GAFF2 overestimates conformational barriers and fails to identify conformer C as a local minimum. \edit{A comparison of the computational cost of the dipole‑autocorrelation MD (Fig.~\ref{fig:of-phe}d) reveals pronounced disparities in performance across the different models. The GAFF2 provides the highest throughput and thus serves as the reference for efficiency. The reported performance of SO3LR reflects that the JAX implementation enables an efficient parallel execution and batching of independent replicas on GPU, leading to a practical slowdown of less than $\sim$10$\times$ relative to GAFF2 in the present protocol. This represents a substantial reduction compared to the larger gap reported in the original SO3LR \cite{Kabylda2024} study, and reflects differences in simulation workflow and parallelization rather than changes in the underlying force field. By contrast, AIMNet2 is roughly $\sim$200$\times$ slower, and MACE-POLAR-1 exhibits a slowdown of approximately $\sim$2000$\times$ relative to GAFF2 in this IR-sampling workflow, partly because the ASE \cite{HjorthLarsen2017} execution strategy used here does not provide the same level of replica-level parallelization and batched GPU throughput. As a consequence, SO3LR uniquely combines accuracy with the computational scalability required for extensive IR MD sampling. This efficiency enables us to provide, in the Supplementary Materials, SO3LR finite‑temperature IR spectra for all 293 molecules in the QVib set (See $IR\_full\_report.pdf$).}\\ 
%AT: 292 or293? [DONE]

\begin{figure}
\centering
\includegraphics[width=\textwidth]{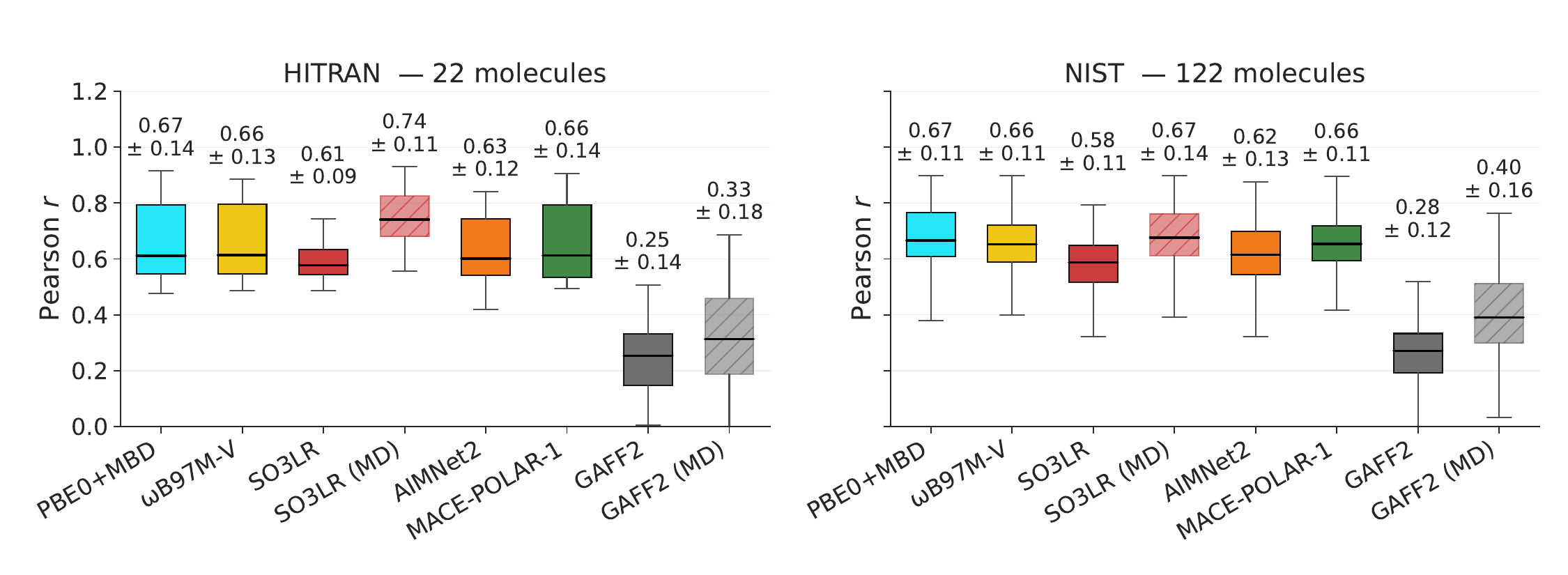}
\caption{\edit{\textbf{Infrared spectral agreement with experiment across all molecules.}
Boxplots show the distribution of Pearson correlation coefficients (r) between predicted and experimental infrared spectra, evaluated at the optimal spectral alignment for each molecule (100 cm\textsuperscript{-1} of allowance). Results are reported separately for gas‑phase HITRAN \cite{HITRAN2024} (22 molecules) and NIST reference database (122 molecules) \cite{NIST35}.
Boxes span the interquartile range (Q1–Q3), the horizontal line marks the median, and whiskers extend to 1.5 × IQR; outliers are not shown. Mean $\pm$ standard deviation of r across molecules is annotated for each method. Harmonic and molecular‑dynamics (MD) variants are reported. Source data are provided as a Source Data file}}
\label{fig:ir_exp}
\vspace{-4.0mm}
\end{figure}

\edit{Ultimately, the relevance of a force field is determined by its ability to reproduce experimentally measured observables. Comparison of computed IR spectra with experimental data for 22 gas‑phase molecules from HITRAN \cite{HITRAN2024} and 122 molecules from the NIST Standard Reference Database 35 \cite{NIST35} reveals a systematic improvement in agreement when finite‑temperature SO3LR molecular dynamics is employed (Fig.~\ref{fig:ir_exp}). This improvement reflects the explicit sampling of anharmonicity, conformational fluctuations, mode coupling, and thermal broadening, all of which contribute to experimental infrared line positions, intensities, and spectral envelopes at finite temperature. Quantitative agreement with experiment, therefore, requires molecular dynamics driven by a force field that is sufficiently accurate to describe the underlying interactions, while remaining computationally efficient enough to enable statistically converged simulations across chemically diverse molecular sets.}\\

\edit{We note that the MD-derived spectra reported here are obtained within a classical-nuclei approximation and therefore do not explicitly include nuclear quantum effects (NQE). Such effects can be relevant, particularly for high-frequency vibrations involving light atoms, and may contribute to residual frequency shifts relative to experiment. In the present work, we account for systematic frequency offsets through the use of a harmonic frequency scaling factor and by allowing a controlled spectral alignment of up to 100~cm$^{-1}$ in the Pearson correlation analysis. For each simulated spectrum, the Pearson correlation with experiment is evaluated over rigid shifts in the range \(-100 \leq \Delta\nu \leq 100\) cm\(^{-1}\) on a common frequency grid, and the maximum correlation is reported together with the corresponding optimal shift. This procedure does not fit individual peaks or alter relative intensities; it preserves the predicted spectral shape and only quantifies the effect of uniform frequency offsets. A rigorous treatment of NQE, for example, through quantum correction factors or path-integral molecular dynamics, is beyond the scope of the present study and will be addressed in future work.}\\

\edit{Taken together, these results support finite-temperature MLFF-driven molecular dynamics as a practical route to experimentally anchored infrared simulations beyond the harmonic approximation. Within the benchmarks examined here, SO3LR provides a favorable compromise between spectral accuracy, explicit treatment of long-range interactions, and computational efficiency, enabling statistically meaningful finite-temperature IR simulations across large and chemically diverse molecular datasets.}\\

\subsection{Stability and Vibrations of Peptides}

\begin{figure}
\centering
\includegraphics[width=1.0\linewidth]{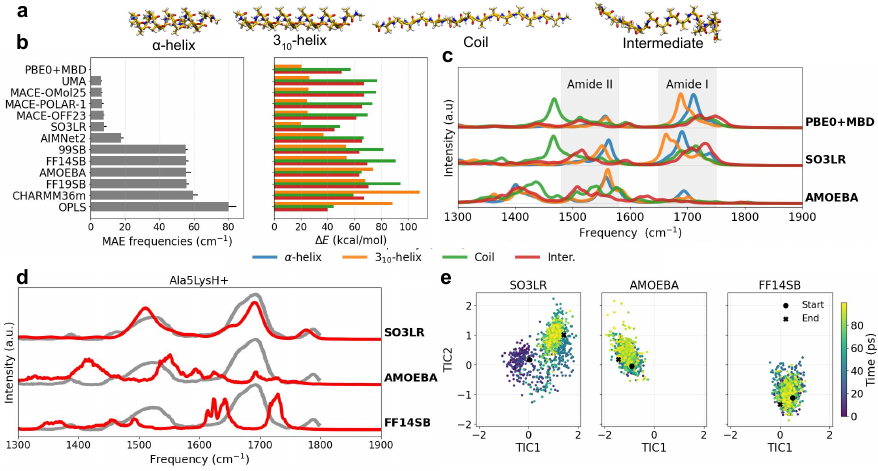}
\caption{\edit{\textbf{Gas-phase peptide secondary-structure analysis.}
\textbf{a} Three-dimensional structures of AceAla\textsubscript{15}NMe for the four conformers considered: canonical $\alpha$-helix, canonical 3\textsubscript{10}-helix, intermediate state, and fully extended coil.
\textbf{b} Frequency mean absolute error (MAE, cm\textsuperscript{-1}, left) and relative potential energies (kcal\,mol\textsuperscript{-1}, right) comparing MLFFs and MMFFs against the PBE0+MBD reference. Frequency MAEs are averaged over the four conformers; error bars denote standard deviations. Relative energies of the 3\textsubscript{10}-helix (orange), coil (green), and intermediate state (red) are shown with respect to the $\alpha$-helix.
\textbf{c} Harmonic infrared spectra of AceAla\textsubscript{15}NMe computed at the PBE0+MBD level and using the SO3LR MLFF and AMOEBA force fields. Conformers are color-coded as follows: $\alpha$-helix (blue), 3\textsubscript{10}-helix (orange), coil (green), and intermediate (red).
\textbf{d} Infrared spectrum of AceAla\textsubscript{5}LysH\textsuperscript{+} obtained with the tested force fields (red), compared with experimental IR reference data (grey) from Ref.~\cite{Rossi2010}. A uniform frequency scaling factor of 0.97 was applied.
\textbf{e} Time-resolved trajectories of Ala$_5$LysH$^{+}$ projected onto a common TICA space~\cite{Perez-Hernandez2013}, obtained from NVE simulations (1 replica, 100\,ps) using different force fields. A single TICA model was trained on pooled backbone $\phi$ and $\psi$ dihedral angles encoded as sine and cosine features. Points are colored by simulation time; circles and crosses denote initial and final conformations, respectively.}
Source data are provided as a Source Data file.}
\label{fig:ala15}
\vspace{-4.0mm}
\end{figure}

We now extend the study from small molecules to larger \edit{peptides}, to investigate whether the quantum mechanically accurate behavior of \edit{MLFFs} observed in small molecules can generalize as molecular size and complexity increase, particularly to systems larger than those in the training dataset. \edit{As an initial} case study, we focus on the AceAla\textsubscript{15}NMe peptide, chosen as a simple but structurally relevant molecule. Its well-characterized folding behavior in vacuum \cite{Millhauser1997, Topol2001} provides a valuable benchmark for evaluating the ability of computational methods to capture secondary structure formation, such as $\alpha$- and 3\textsubscript{10}-helices. Once formed, helices can dynamically interconvert between canonical $\alpha$- and 3\textsubscript{10}-helical structures. The folding behavior has been successfully reproduced by DFT-based MD simulations \cite{Tkatchenko2011}, many-body potentials \cite{Zhou2025}, as well as by pretrained MLFFs such as GEMS \cite{Unke2024}, MACE-off23 \cite{Kovacs2025}, and SO3LR \cite{Kabylda2024}; here we validate the folding (not captured by MMFF, See Sup. Fig. A5) to then investigate dynamics in detail.\\

NMA was performed on four representative snapshots from the folding trajectory (Fig.~\ref{fig:ala15}a), with backbone heavy atoms constrained during minimization to prevent structural deviations: (1) a completely unfolded state (coil), (2) an intermediate conformation, (3) a canonical $\alpha$-helix, and (4) a canonical $3_{10}$-helix revealing that the \edit{MLFFs} tested reproduces the vibrational modes with high fidelity (Fig. \ref{fig:ala15}b). The six MMFFs tested show significantly lower agreement with DFT. \edit{The difference MD behaviour between SO3LR and MMFFs is explained by the enthalpic stability differences between the canonical $\alpha$ and 3\textsubscript{10}. MMFFs typically did not capture this equilibrium well, overestimating energy barriers. These results confirm explicit DFT+vdW calculations~ \cite{Tkatchenko2011}.}\\ 

\edit{Amide-I (1600-1700 cm\textsuperscript{-1}) and amide-II vibrational (1480-1580 cm\textsuperscript{-1}) bands are widely used in infrared spectroscopy as sensitive probes of protein and peptide secondary structure \cite{Tintor2024}. Using the AceAla\textsubscript{15}NMe model system, the corresponding IR spectrum was computed and benchmarked against a PBE0+MBD reference (Fig. \ref{fig:ala15}c and See. Sup. Fig. A6). The SO3LR and MACE‑POLAR‑1 models show very good agreement with the \textit{ab initio} spectrum, correctly reproducing both the position and the relative intensity of the amide bands and enabling a clear discrimination of the underlying secondary structure. MMFFs fixed charge partitions fail to capture these spectral signatures. AMOEBA partially alleviates this limitation through the inclusion of explicit multipolar electrostatics and inducible dipoles, but their description of subtle vibrational features remains qualitatively incomplete.}\\ 

\edit{To further assess the predictive power of the models under experimentally relevant conditions, we turn to a direct comparison with gas‑phase IR measurements. AceAla\textsubscript{5,10,15}LysH+ was characterized in the gas phase, and IR spectra were obtained from finite‑temperature MD simulations, enabling a direct connection between conformational ensembles and vibrational response. In the work of Rossi \textit{et al.} \cite{Rossi2010}, a combination of gas‑phase IR experiments and \emph{ab initio} MD established that AceAla\textsubscript{5}LysH+ populates a mixed ensemble of helical and non‑helical conformations at room temperature, whereas helical structures become predominant for AceAla\textsubscript{10,15}LysH+. Consistent with these findings, the SO3LR model yields IR spectra in excellent agreement with both experiment and \emph{ab initio} molecular dynamics based on PBE+vdW \cite{Rossi2010}, accurately reproducing band positions and relative intensities across all peptide lengths  (Fig. \ref{fig:ala15}d and see Sup. Fig. A7-9). Importantly, the resulting spectra emerge from the weighted contribution of coexisting secondary‑structure motifs sampled along the dynamics, underscoring the need for an environment‑dependent and structurally adaptive description of the potential energy surface (Fig. \ref{fig:ala15}e and see Sup. Fig. A10-A12). MMFFs, including polarizable variants, lack this representability and consequently fail to reproduce the experimental spectral fingerprints arising from subtle secondary‑structure heterogeneity.}\\

\edit{The ability to reproduce vibrational spectra and conformational dynamics at MLFF cost directly impacts biomolecular modelling. Infrared spectra of peptides and proteins are not simple fingerprints of local force constants; amide-I and amide-II bands encode ensemble-averaged secondary-structure populations, hydrogen-bonding networks, electrostatic environments, and thermal fluctuations \cite{Barth2007, Tintor2024}. Consequently, an incorrect energetic landscape produces an incorrect spectrum. This is observed here for conventional MMFFs, including polarizable variants, which do not consistently reproduce the DFT-level balance between competing peptide conformations and therefore fail to recover the experimental amide-band signatures reported for gas-phase peptides \cite{Rossi2010, Tkatchenko2011}. MLFFs such as SO3LR overcome this limitation by retaining near-DFT fidelity for both conformational energetics and vibrational response at a cost compatible with finite-temperature sampling. Agreement with experimental spectroscopy therefore provides more than a spectral benchmark: it validates the predicted biomolecular dynamics and enables atomistic interpretation of peptide and protein structure, flexibility, and interactions beyond the reach of conventional force-field approaches.}

\subsection{Environment Effects on Vibrations in a Multimeric Protein}

\begin{figure}
\centering
\includegraphics[width=\textwidth]{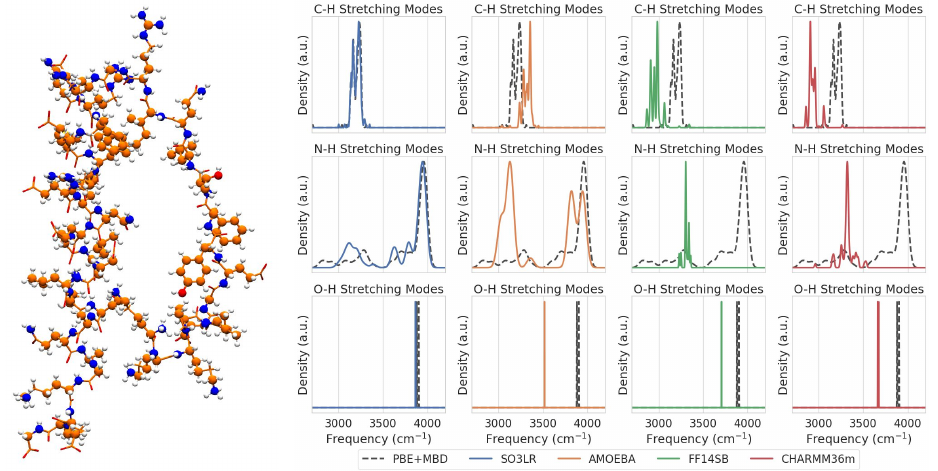}
\caption{\textbf{High frequency stretching modes in p53 monomer}. Normalized spectral profiles of C-H, N-H and O-H modes from 2700 to 4200 cm\textsuperscript{-1} in p53 monomer (PDBid: 1SAE) \cite{Clore1995}. Different computational methods are shown: PBE+MBD (dotted black), SO3LR (blue), AMOEBA (orange), FF14SB (green) and CHARMM36m (red). Source data are provided as a Source Data File.}
\label{fig:p53corr}
\vspace{-4.0mm}
\end{figure}

\begin{figure*}%[tbhp]
\centering
\includegraphics[width=\textwidth]{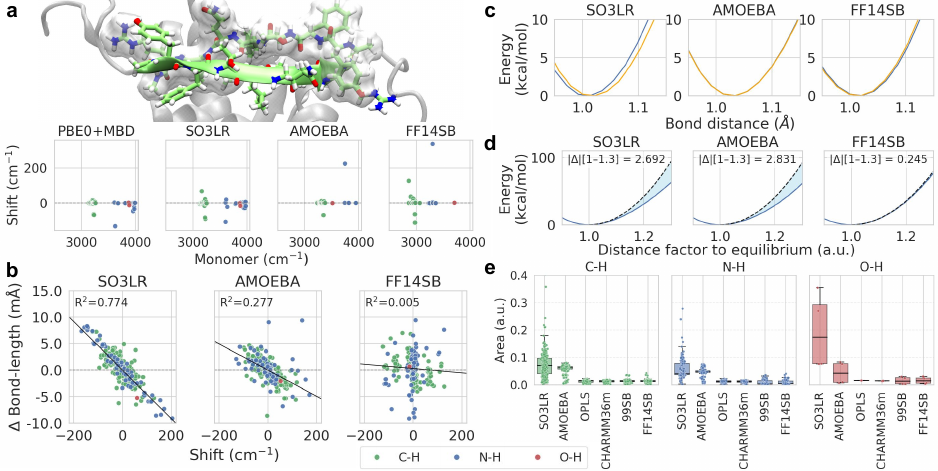}
\caption{\textbf{Environmental effects on stretching modes in p53 multimeric protein.} C–H (green), N–H (blue), and O–H (red) stretching modes in the 2700-4200~cm\textsuperscript{-1} region are studied under different conditions.
\textbf{a} Vibrational shifts between dimer and monomer configurations (x‑axis) for the capped fragment of residues 327–333 (green), computed using PBE0+MBD, SO3LR, AMOEBA, and FF14SB.
\textbf{b} Vibrational shifts plotted against bond‑length differences between solvated tetramer and monomer configurations, including the linear correlation $R^2$
\textbf{c} Potential energy profiles for bond‑length perturbations around equilibrium for the N–H stretching mode of the Ile332 backbone, shown for the monomer (blue) and tetramer (orange).
\textbf{d} Energy comparison for the same atoms as a function of scaled bond distance (0.9–1.3× equilibrium) in the monomer. The shaded region represents the area predicted by a fully harmonic potential from 1.0 to 1.3× the equilibrium distance (dotted).
\textbf{e} Normalized area distributions (scaled to 1 at maximum) as an anharmonicity measure comparing SO3LR and MMFFs, grouped by bond type.
In scatter plots, each point corresponds to the most similar mode matched by the atom index. Source data are provided as a Source Data File.}
\label{fig:p53_wat}
\end{figure*}

We now explore a multimeric protein system to evaluate how SO3LR and MMFFs methodologies capture intermolecular and intramolecular interactions within varying environments in complex biomolecular systems. We investigated the tetrameric \edit{oligomerization} domain of p53 (PDB ID: 1SAE), a well-characterized protein known for its structural stability in the tetrameric form. In contrast, its monomeric state is reported to be significantly less stable and prone to unfolding or misfolding \cite{Clore1995, Gunasekaran2004}. A key question is whether MLFFs trained on small molecules can accurately extrapolate and reproduce the behaviour of large biomolecular systems, a bottom-up generalisation challenge. To address this, we performed NMA of the p53 monomer extracted from crystal structures \cite{Clore1995} at the PBE+MBD level of theory and compared the resulting vibrational modes with those predicted by SO3LR and MMFF models, since higher-level calculations (PBE0+MBD) were not feasible for this system size.\\

The MLFF approach again shows promising fidelity with much lower deviations compared to MMFFs. Although SO3LR was not trained on PBE+MBD data, it nevertheless reproduces the reference with notably better accuracy than MMFFs, including in the high‑frequency region of the vibrational density of states where some discrepancies remain (see Sup. Fig. A13). When vibrational modes are resolved by chemical identity, SO3LR captures the characteristic C–H, N–H, and O–H stretching distributions far more faithfully than MMFFs (Fig.~\ref{fig:p53corr}). An intermediate case is the AMOEBA polarizable FF, which captures these stretching modes more accurately than classical MMFFs. These modes are particularly sensitive to fine electronic structure details and anharmonic contributions that classical MMFFs struggle to represent.\\

We next examined protein–protein interactions specifically focusing on the $\beta$-sheet region (residues \edit{3}27–333) where dimerization of the p53 protein occurs (Fig.~\ref{fig:p53_wat}a, Sup. Fig. A14). The vibrational profile obtained using SO3LR closely matches the reference data from PBE0+MBD, indicating that SO3LR effectively captures the subtle interplay of intra- and intermolecular forces that dominate the dynamics of the dimer interface. In contrast, MMFFs yield significantly different profiles, often failing to reproduce the nuanced packing effects observed in the DFT-level calculations.\\

In aqueous solution, solvent‑dependent shifts arise from a balance of hydrogen‑bonding, local electrostatic fields, steric hindrance, and mode coupling. Hydrogen bonding typically elongates and softens O–H and N–H bonds, producing red shifts in their stretching frequencies \cite{Brunig2022a}. Blue shifts can occur in scaffolds exhibiting improper H bonds or when strong local electric fields stiffen the X–H bond—effects showing that electrostatic charge distribution and hydrogen‑bond geometry can either red‑ or blue‑shift stretching modes depending on the orientation and nature of the interaction \cite{Selvam2010, Kirsh2024}. \\

A comparative analysis of solvation effects in tyrosine, arginine, and leucine, with PBE0+MBD as the reference, shows that SO3LR reliably reproduces solvent-induced vibrational shifts, whereas MMFFs fail to capture these effects (See Supp. Appendix B). For tyrosine, the O–H stretch undergoes a pronounced red shift in both DFT and SO3LR, due to hydrogen bonding of water with the tyrosine hydroxy group, while MMFFs predict minimal change. The N–H groups of arginine exhibit both red and blue solvent-dependent displacements, which are captured by SO3LR but missed by MMFFs. The C–H stretching modes in all molecules tested show small blue shifts in both SO3LR and DFT, with larger shifts observed in arginine due to the stronger electric fields of its charged groups, while MMFFs yield unrealistic deviations.\\

Building on these findings, we next examine the p53 protein in a solvated environment, comparing monomeric and tetrameric states to uncover the mechanistic origins of the observed vibrational shifts in a biologically relevant system (see Sup. Fig. A15). Our goal is to determine how these shifts arise from structural adjustments—such as elongation or shortening of X–H bonds—and how they are influenced by the harmonic or anharmonic nature of the vibrational potential. To this end, we compare the N–H, C–H, and O–H stretching frequency shifts with the corresponding bond-length differences between monomer and tetramer, focusing on variations within the range of –10 to +10 m\AA~(Fig. \ref{fig:p53_wat}b and Sup. Fig. A16). Remarkably, SO3LR exhibits a clear linear correlation between bond-length changes and vibrational shifts, indicating that the model captures the underlying physical relationship. In contrast, AMOEBA shows only a weak linear trend, while MMFFs fail to reproduce any meaningful correlation.\\

To illustrate these effects, we examine the N–H stretching mode of the Ile332 backbone amide, which stabilizes the dimerization H‑bond with the partner monomer. SO3LR predicts a red shift of 145 cm\textsuperscript{-1}, whereas AMOEBA and FF14SB yield blue shifts of 15 and 45cm\textsuperscript{-1}, respectively. Inspection of the potential energy profile for bond‑length perturbations from 0.9 to 1.3 (Fig. \ref{fig:p53_wat}c) shows that SO3LR uniquely alters both the equilibrium position and the potential shape due to its environment awareness, while the fixed functional forms of the MMFFs produce nearly identical profiles. Both SO3LR and AMOEBA reproduce an anharmonic bond‑stretch profile (Fig.~\ref{fig:p53_wat}d), but the origin of this anharmonicity differs: While SO3LR learns it directly from the QM training data, AMOEBA introduces an analytic anharmonic bond potential \cite{Shi2013}. This explains the trend in Fig.\ref{fig:p53_wat}e, where the normalized area between the fully harmonic and computed potentials is compared: the flexible, environment‑responsive SO3LR potential captures DFT‑like behavior; AMOEBA’s built‑in anharmonicity improves its shifts relative to non‑polarizable MMFFs, though its fixed form still limits adaptability.\\

MMFFs systematically fail to reproduce these behaviours because they lack explicit polarization, many‑body induction, anharmonicity, and environment‑sensitive hydrogen‑bond descriptions. Their harmonic bond potentials and fixed charges cannot capture the solvent‑dependent, cooperative stabilization patterns observed in strongly hydrogen‑bonded systems. Since bond‑stretch anharmonicity is a primary driver of frequency shifts, harmonic force fields predict little change upon stretching; mode coupling can shift frequencies, but this mechanism alone yields incorrect behavior.  Accurately reproducing these shifts requires a quantum‑mechanical treatment of the dynamical behavior of electrons and anharmonicity. An important exception among MMFFs is AMOEBA, which incorporates explicit polarization, multipole electrostatics, and anharmonic bond terms. These features allow AMOEBA to capture hydrogen‑bond cooperativity and vibrational shifts more faithfully than non-polarizable MMFFs \edit{\cite{Kirsh2026}}. However, despite its improved physical realism, AMOEBA still relies on fixed functional forms and parameter sets that do not fully capture the changing chemical environments.

\subsection{Final Remarks}
MMFFs remain widely used in biomolecular simulations, yet their reliance on fixed functional forms and empirical fitting limits their ability to capture complex vibrational and environmental effects. Although the AMOEBA polarizable MMFF performs slightly better than non-polarizable MMFFs by incorporating explicit polarization and anharmonic terms, its fixed functional forms still \edit{restrict the accurate description of solvent effects, protein–protein interactions, and conformational energetics across diverse environments. As a result, inaccuracies in the underlying potential energy landscape can propagate into errors in predicted dynamics and spectroscopic observables.}\\ 

Our results highlight the transformative potential of MLFFs, \edit{exemplified} by SO3LR, \edit{can bridge this gap by learning interatomic interactions directly from quantum‑mechanical reference data. Across the systems investigated, MLFFs reproduce vibrational frequencies, infrared intensities, and finite‑temperature spectral signatures in close agreement with both \emph{ab initio} calculations and experimental measurements. The ability to match experimental spectroscopy is particularly important, as vibrational observables provide a stringent and independent validation of the underlying force field, probing not only local bonding but also long‑range coupling, polarization, and conformational heterogeneity. Agreement with experiment therefore establishes confidence that the predicted dynamics and energetics are physically meaningful, rather than artefacts of parametrization.}\\

\edit{From a biomolecular perspective, this capability is critical. Infrared spectra of peptides and proteins arise from thermally averaged ensembles of interconverting conformations, and their interpretation requires a force field that accurately couples structural dynamics to vibrational response while remaining computationally tractable. MLFFs such as SO3LR achieve this balance, enabling finite‑temperature simulations with near‑DFT fidelity at a cost compatible with extended biomolecular sampling. By providing quantitative agreement with experimental spectroscopy alongside accurate dynamics, MLFF‑based simulations offer a validated and practical framework for studying biomolecular structure, flexibility, and interactions, extending the predictive scope of molecular simulations beyond what is accessible with conventional force‑field approaches.}

\section{Methods}

\subsection{Density-Functional Theory calculations}
DFT calculations were performed using the FHI-aims software package (version \edit{250320}) \cite{Blum2009}, employing the PBE0 hybrid exchange-correlation functional \cite{Adamo1999} combined with the Many-Body Dispersion (MBD) correction scheme to account for long-range van der Waals interactions \cite{Tkatchenko2012, Ambrosetti2014} \edit{and using ORCA 6.0.0 \cite{ORCA6} with the $\omega$B97M-V fucntional}. For geometry optimizations and energy evaluations, \edit{FHI-aims} "tight" numerical settings \edit{and ORCA def2-TZVPD bases} were applied to small organic molecules, "intermediate" settings were used for the larger AceAla\textsubscript{15}NMe peptide and the Ace-$\beta$-sheet-NMe dimer and monomer (residues 327 to 333) of p53 system to balance accuracy and computational cost. Due to the high computational expense, the p53 monomer was computed at the PBE+MBD \cite{Perdew1996} level of theory using “light” settings.
\subsection{Molecular dynamics} \label{molecular_dynamics}
Classical MMFFs dynamics were performed using OpenMM 8.1.1 \cite{Eastman2024} with the Langevin middle integrator \cite{Zhang2019}. In total, seven widely used classical force fields were evaluated: 99SB-ILDN \cite{Lindorff-Larsen2010}, FF14SB \cite{Maier2015}, and FF19SB \cite{Tian2020} from the AMBER family; CHARMM36m \cite{Huang2017}; OPLS-AA \cite{Robertson2015}, and the polarizable AMOEBA force field \cite{Shi2013, Zhang2018}.
SO3LR dynamics were performed using the jax-md \cite{jaxmd2020} \edit{and a long-range cut-off of 1000 \AA. MACE-POLAR-1 and AIMNet2 MD were performed using ASE \cite{HjorthLarsen2017}. The Langevin integrator was used for MD with MLFFs}.

\subsection{Normal Mode Analysis}
Each molecule was geometry-optimized using different methods depending on the level of theory. For classical MMFF, the L-BFGS algorithm implemented in OpenMM was employed. For MLFFs, optimization was performed using ASE~\cite{HjorthLarsen2017}. DFT calculations were carried out using the BFGS algorithm in FHI-aims \edit{and ORCA. The convergence criterion was set to $5 \times 10^{-3}$~eV/\AA~in all methods.} \\
\edit{For each small molecule, conformers were obtained from the PubChem database~\cite{Bolton2011} and subsequently optimized at the PBE0+MBD and $\omega$B97M-V levels of theory. To minimize deviations in the NMA, the PBE0+MBD-optimized structures were used as the starting geometries for the force fields.}\\
For the AceAla\textsubscript{15}NMe peptide, during geometry optimization, the backbone carbon and nitrogen atoms were constrained to preserve the backbone conformation.
The p53 monomer and tetramer structures used in calculations were taken directly from the reference structure (PDB ID: 1SAE)~\cite{Clore1995}. In solvated systems, molecules were optimised with a convergence criteria of $5 \times 10^{-2}$~eV/\AA~in the case of SO3LR.\\
Numerical Hessians were computed using finite displacements: 0.005~\AA\ for classical MMFF, and 0.01~\AA\ for both MLFF and DFT. In the case of multimeric proteins, only the protein atoms were displaced during Hessian construction, while solvent atoms were included in the force evaluations. Normal mode analysis was then performed to extract vibrational eigenvectors and frequencies from the Hessians.

\subsection{Infrared spectrum}\label{IR}
20 independent 100 ps NVE trajectories with a 0.2 fs/step were used after 10 ps NVT equilibration. 
The instantaneous dipole moment of the system was calculated at each step. \edit{For fixed-charge MMFFs models and for partial‑charge‑predicting MLFF (SO3LR and AIMNet) models}, the dipole moment vector, $\vec{\mu}(t)$, was obtained using the following equation:
\begin{equation}%[tbhp] 
\vec{\mu}(t) = \sum_i q_i \vec{r}_i(t),
\end{equation}
where $q_i$ is the charge of the $i$-th atom and $\vec{r}_i(t)$ is its position at time $t$. \edit{In MACE-POLAR-1, the molecular dipole moment was taken directly from model prediction. In the case of the polarizable AMOEBA force field, the total dipole moment was also used.}
The dipole moment autocorrelation function (ACF) was computed from the time series of dipole moments. The ACF, $C_{\mu}(t)$, is defined as:
\begin{equation}%[tbhp] 
C_{\mu}(t) = \langle \vec{\mu}(0) \cdot \vec{\mu}(t) \rangle.
\end{equation}
The IR spectrum was obtained by performing a Fourier transform of the dipole moment ACF. The power spectrum, $I(\omega)$, is given by:
\begin{equation}%[tbhp] 
I(\omega) = \int_{-\infty}^{\infty} C_{\mu}(t) e^{-i\omega t} dt,
\end{equation}
where $\omega$ is the angular frequency. The final spectrum is obtained by averaging over all trajectories.\\
\edit{DFT level IR spectra were computed within the harmonic approximation at the PBE0+MBD and $\omega$B97M-V levels of theory. Equilibrium structures were optimized at each level, followed by evaluation of the (mass-weighted) Hessian by finite differences to obtain harmonic normal-mode frequencies, $\omega_k$, and eigenvectors. IR intensities were computed from the derivative of the molecular dipole moment with respect to the normal coordinate $Q_k$,
\begin{equation}
I_k \propto \left|\frac{\partial \vec{\mu}}{\partial Q_k}\right|^2,
\end{equation}\\
This protocol was also followed with MLFFs and MMFFs for Harmonic approximation IR. A uniform frequency scaling factor of 0.97 was applied to all computed IR spectra, unless stated otherwise.}\\

\subsection{\edit{Cross correlation metrics}}
\label{cross_corr}
\edit{To quantify agreement between method and reference IR spectra in a way that is robust to small uniform frequency offsets, we computed a spectral cross-correlation similarity. The harmonic approximation spectra were first broadened with Lorentzian line shapes of $\mathrm{FWHM}= 15$~cm$^{-1}$ on a common frequency grid spanning $[500, 4000]$~cm$^{-1}$ at $\Delta\nu = 1$~cm$^{-1}$ resolution, $\tilde{I}_\text{ref}(\nu)$ and $\tilde{I}_\text{method}(\nu)$, and each was normalised to unit $L^2$ norm. The shift-dependent cross-correlation
\begin{equation}
\rho(\Delta) \;=\; \int \tilde{I}_\text{ref}(\nu)\, \tilde{I}_\text{method}(\nu - \Delta)\, d\nu
\end{equation}
was evaluated for $\Delta\nu = 1$~cm$^{-1}$ rigid shifts $\Delta \in [-50, +50]$~cm$^{-1}$. The reported \emph{Corr} value is the maximum cross-correlation,
\begin{equation}
\mathrm{Corr} \;=\; \max_{|\Delta| \le 50\,\text{cm}^{-1}} \rho(\Delta) \;\in\; [0, 1],
\end{equation}
which equals 1 for spectra that coincide up to a rigid translation and decays towards 0 as their band positions and intensities diverge. The associated \emph{Shift},
\begin{equation}
\mathrm{Shift} \;=\; \operatorname*{arg\,max}_{|\Delta| \le 50\,\mathrm{cm}^{-1}} \rho(\Delta),
\end{equation}
is the rigid frequency offset (in cm$^{-1}$) that maximises overlap with the reference: a positive value indicates that the method's bands are systematically blue-shifted with respect to the reference, a negative value that they are red-shifted. Together, Corr and Shift separate band-shape agreement from systematic frequency miscalibration.}

\subsection{Potential energy surface}
Two dihedral angles of L-$o$F-phenylalanine+H\textsuperscript{+}, $\phi_1$ (N–C$_\alpha$–C$_\beta$–C$_\gamma$) and $\phi_2$ (H$_{\beta^1}$–C$_\beta$–C$_\gamma$–C$_{\sigma^2}$), were selected for torsional analysis. A 10° rotational window was applied, using the most stable conformation as the starting point. For each position, fixed-dihedral minimization was performed with \edit{MLFFs} and GAFF2. In GAFF2, an external dihedral constraint of 50 kcal·mol$^{-1}$ was imposed. For DFT, constrained relaxation was first carried out with ORCA at the PBE0  \edit{and $\omega$B97M-V level}, followed by FHI-aims at PBE0+MBD \edit{(from PBE0 structures)}, where dihedral atom positions were fixed due to the lack of direct dihedral constraints in FHI-aims.

\subsection{Molecular Visualization}
Three-dimensional molecular structures were visualized using VMD version 1.9.4a57 \cite{HUMP96}.

\section{Data availability}
\edit{ The QVib benchmark dataset, including all computed harmonic and finite-temperature infrared spectra, normal mode analyses, potential energy surfaces, peptide and protein vibrational data, source data
  underlying all figures, and figure-generation scripts, is deposited on Zenodo (DOI: 10.5281/zenodo.20315304). Experimental infrared spectra from the HITRAN 2024 database (\url{https://hitran.org}) and NIST
  Standard Reference Database 35 are available from their respective repositories; licensing restrictions prevent redistribution. Reference QM7-X data are available from Zenodo (DOI: 10.5281/zenodo.4288677). The
   p53 oligomerization domain structure is available from the Protein Data Bank under accession code 1SAE. Small-molecule conformers were obtained from the PubChem database
  (\url{https://pubchem.ncbi.nlm.nih.gov}).
}.

\bibliography{references.bib}

\section{Acknowledgements}
The authors express their gratitude to Ariadni Boziki for the technical support on the IR spectrum and numerical Hessian, and Adil Kabylda and Tobias Henkes for their support in MD simulations.
S.S.D, F.N.B. and A.T. acknowledge support from the Luxembourg National Research Fund under grant FNR-CORE MBD-in-BMD (18093472) and the European Research Council under ERC-AdG grant FITMOL (101054629). M.G. from the European Union’s Horizon Europe Marie Skłodowska-Curie Actions (MSCA) Postdoctoral Fellowship (101202630). K.H and J.T.B. acknowledge support from the FNR (C20/MS/14588607).
The computations were performed on the Luxembourg national supercomputer MeluXina and in the Uni.lu.
\section{Author Contributions}
S.S.D. and A.T. designed the research. S.S.D. performed all simulations, carried out data analysis, and wrote the manuscript with input from all the authors. M.G., K.H and F.N.B. provided technical support and scientific discussion throughout the study.
J.T.B. and A.T. contributed to the interpretation of the results and the revision of the manuscript. All authors have read and approved the final version of the manuscript. A.T. supervised the project and guided the scientific discussions.
\section{Competing interests}
The authors declare no competing interests.
\section{Additional Information}
\subsection{Supplementary information}
The following files are available free of charge.
\begin{itemize}
  \item SI.pdf: 
  \begin{itemize}
      \item Appendix A: Supplementary Figures A1-A\edit{16}
      \item Appendix B: Amino Acids under Solvent Effects (Figures B1-B3)
  \end{itemize}
\end{itemize}
\subsection{Correspondence}
Correspondence and requests for materials should be addressed to Alexandre Tkatchenko.
\end{document}

\title[Supplementary information: Quantum-Accurate Vibrational and Conformational Dynamics from Small Molecules to Proteins with Machine-Learned Force Fields]{Supplementary information: Quantum-Accurate Vibrational and Conformational Dynamics from Small Molecules to Proteins with Machine-Learned Force Fields}
\author[1]{\fnm{Sergio} \sur{Suárez-Dou}}
\author[1]{\fnm{Miguel} \sur{Gallegos}}
\author[1]{\fnm{Kyunghoon} \sur{Han}}
\author[1]{\fnm{Florian N.} \sur{Brünig}}
\author[1]{\fnm{Joshua T.} \sur{Berryman}}
\author*[1]{\fnm{Alexandre} \sur{Tkatchenko}}
\affil[1]{Department of Physics and Materials Science, University of Luxembourg, L-1511 Luxembourg City, Luxembourg}
\maketitle

\newpage
\begin{appendices}
\numbered
\section{Supplementary Figures A1-16}

\begin{figure*}[b]
\centering
\includegraphics[width=\linewidth]{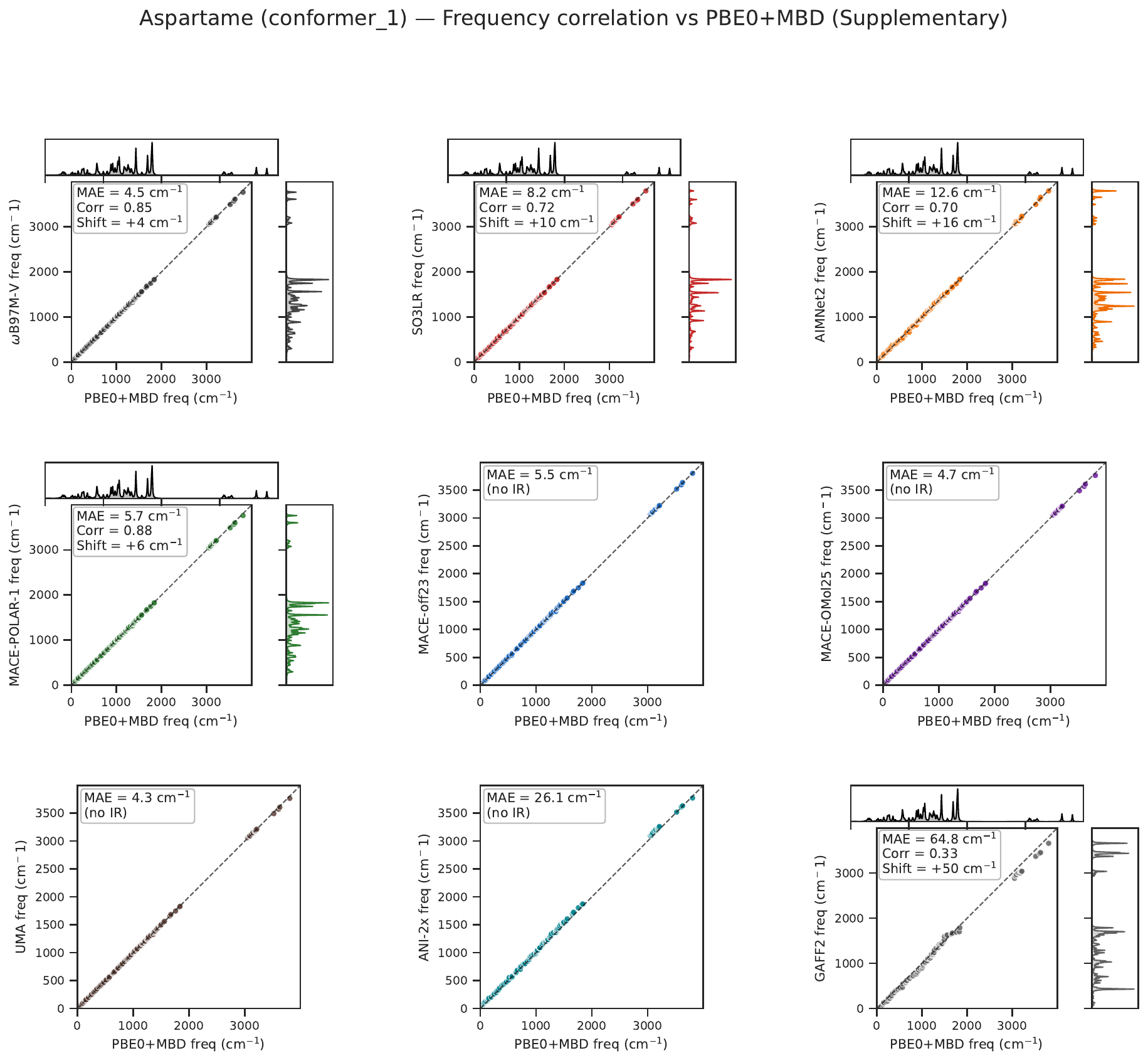}
\caption{Per-method frequency correlation against PBE0+MBD for aspartame (conformer 1). For methods that provide IR intensities, the top and right marginals show Lorentzian-broadened IR spectra (FWHM = 15 cm\textsuperscript{-1}) of the reference and the method, respectively; methods without IR intensities (ANI-2x, MACE-off23, MACE-OMol25, UMA) display only the frequency scatter. Each panel reports the frequency MAE and, where applicable, the IR spectral cross-correlation.}
\label{sfig:so3lr_cover}
\end{figure*}

\begin{figure*}[b]
\centering
\includegraphics[width=\linewidth]{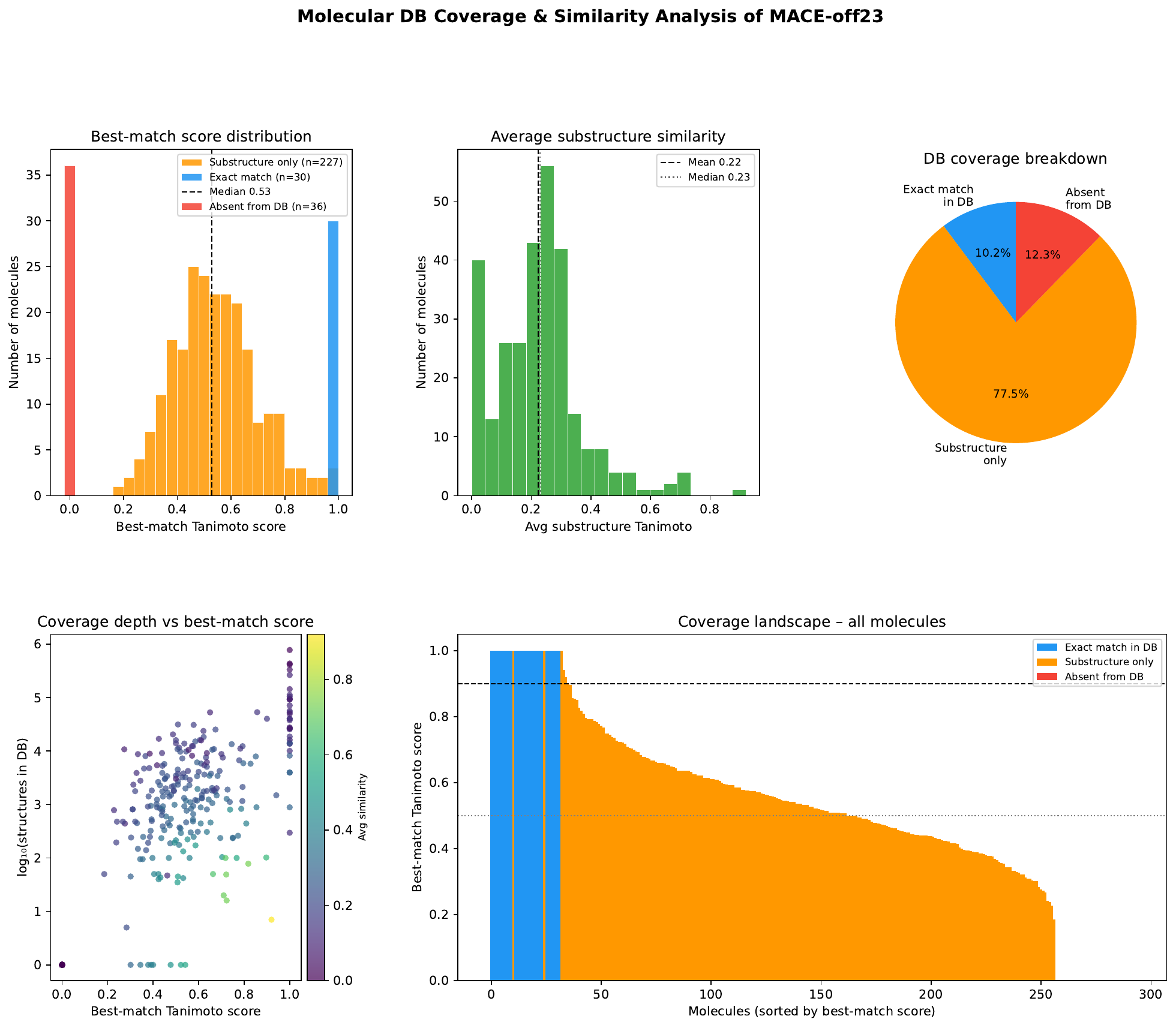}
\caption{Database coverage and chemical similarity analysis for the MACE‑off23 training set relative to the benchmark small‑molecule dataset.
All molecules in the benchmark were queried against the MACE‑off23 database, and coverage was quantified in terms of exact matches, substructure-only matches, and complete absence. (Top left) Distribution of best‑match Tanimoto similarity scores, stratified by match type (exact, substructure only, and absent). (Top middle) Distribution of average substructure similarity per molecule. (Top right) Overall coverage breakdown showing the fraction of molecules with exact matches, substructure-only matches, or no representation in the database. (Bottom left) Depth of coverage, shown as the number of database structures containing the target or a substructure (log scale) versus best‑match similarity, colored by average substructure similarity. (Bottom right) Coverage landscape across all molecules, sorted by best‑match similarity score, highlighting molecules that are well represented, partially represented, or entirely absent from the training data.}
\label{sfig:mace_cover}
\end{figure*}

\begin{figure*}[b]
\centering
\includegraphics[width=\linewidth]{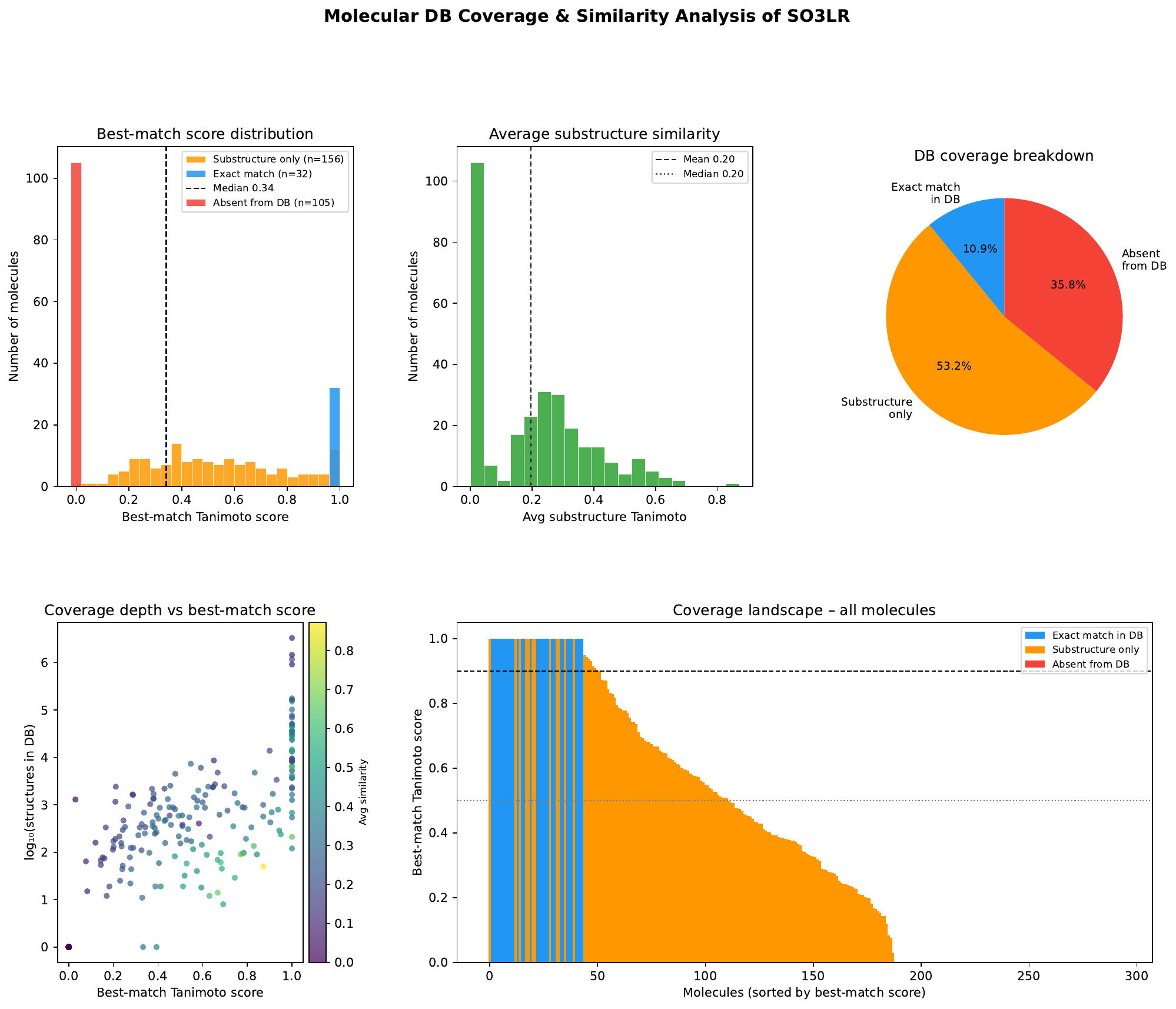}
\caption{Database coverage and chemical similarity analysis for the SO3LR training set relative to the benchmark small‑molecule dataset.
All molecules in the benchmark were queried against the SO3LR database, and coverage was quantified in terms of exact matches, substructure-only matches, and complete absence. (Top left) Distribution of best‑match Tanimoto similarity scores, stratified by match type (exact, substructure only, and absent). (Top middle) Distribution of average substructure similarity per molecule. (Top right) Overall coverage breakdown showing the fraction of molecules with exact matches, substructure-only matches, or no representation in the database. (Bottom left) Depth of coverage, shown as the number of database structures containing the target or a substructure (log scale) versus best‑match similarity, colored by average substructure similarity. (Bottom right) Coverage landscape across all molecules, sorted by best‑match similarity score, highlighting molecules that are well represented, partially represented, or entirely absent from the training data.}
\label{sfig:so3lr_cover}
\end{figure*}

\begin{figure*}[b]
\centering
\includegraphics[width=\linewidth]{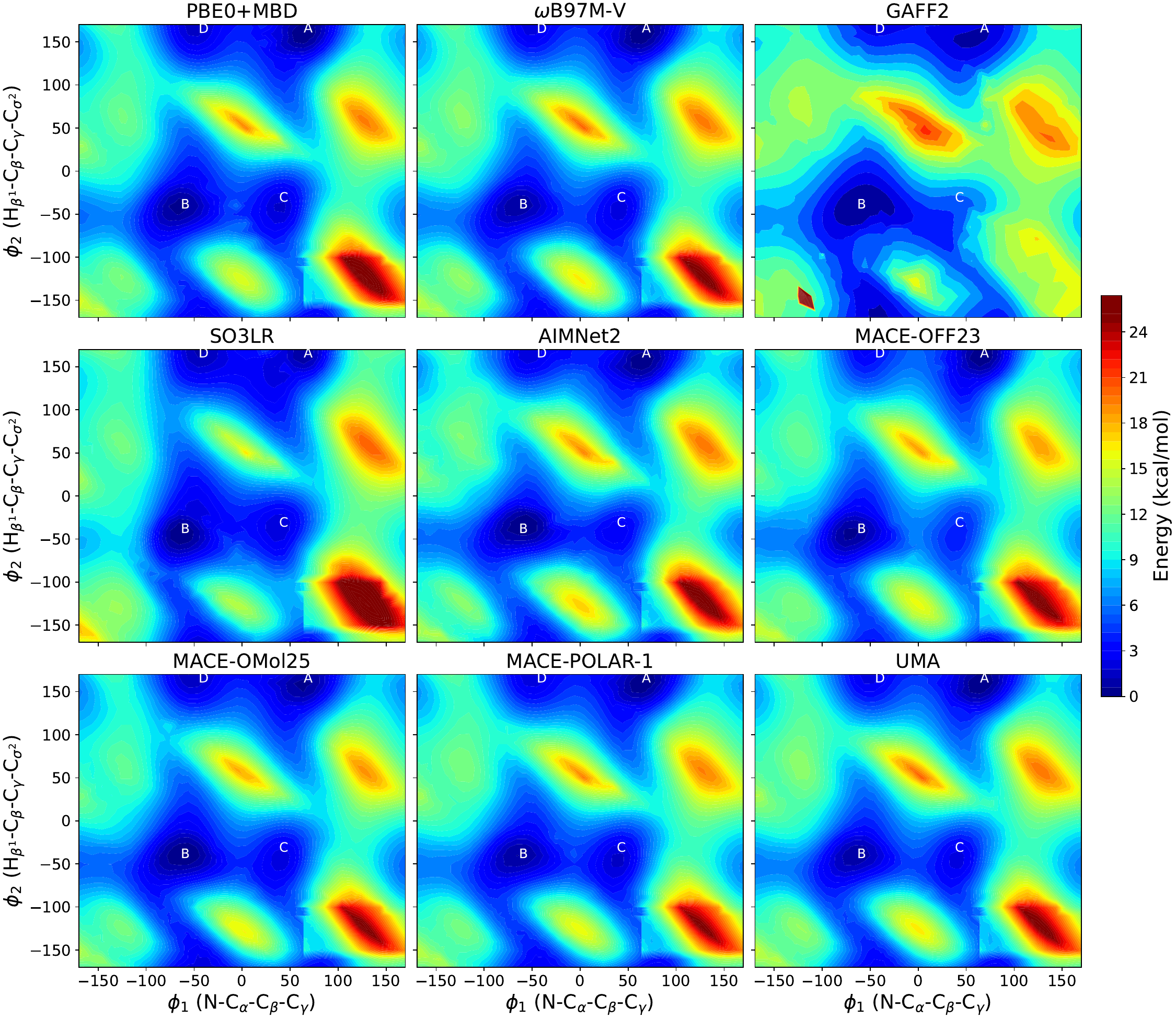}
\caption{Potential energy surface of oF-Phe molecule mapped as a function of $\phi_1$ (N–C$_\alpha$–C$_\beta$–C$_\gamma$) and $\phi_2$ (H$_{\beta^1}$–C$_\beta$–C$_\gamma$–C$_{\sigma^2}$) dihedral angles. Letters indicate conformers defined in \cite{Safferthal2023}.}
\label{sfig:pes}
\end{figure*}

\begin{figure*}[b]
\centering
\includegraphics[width=\linewidth]{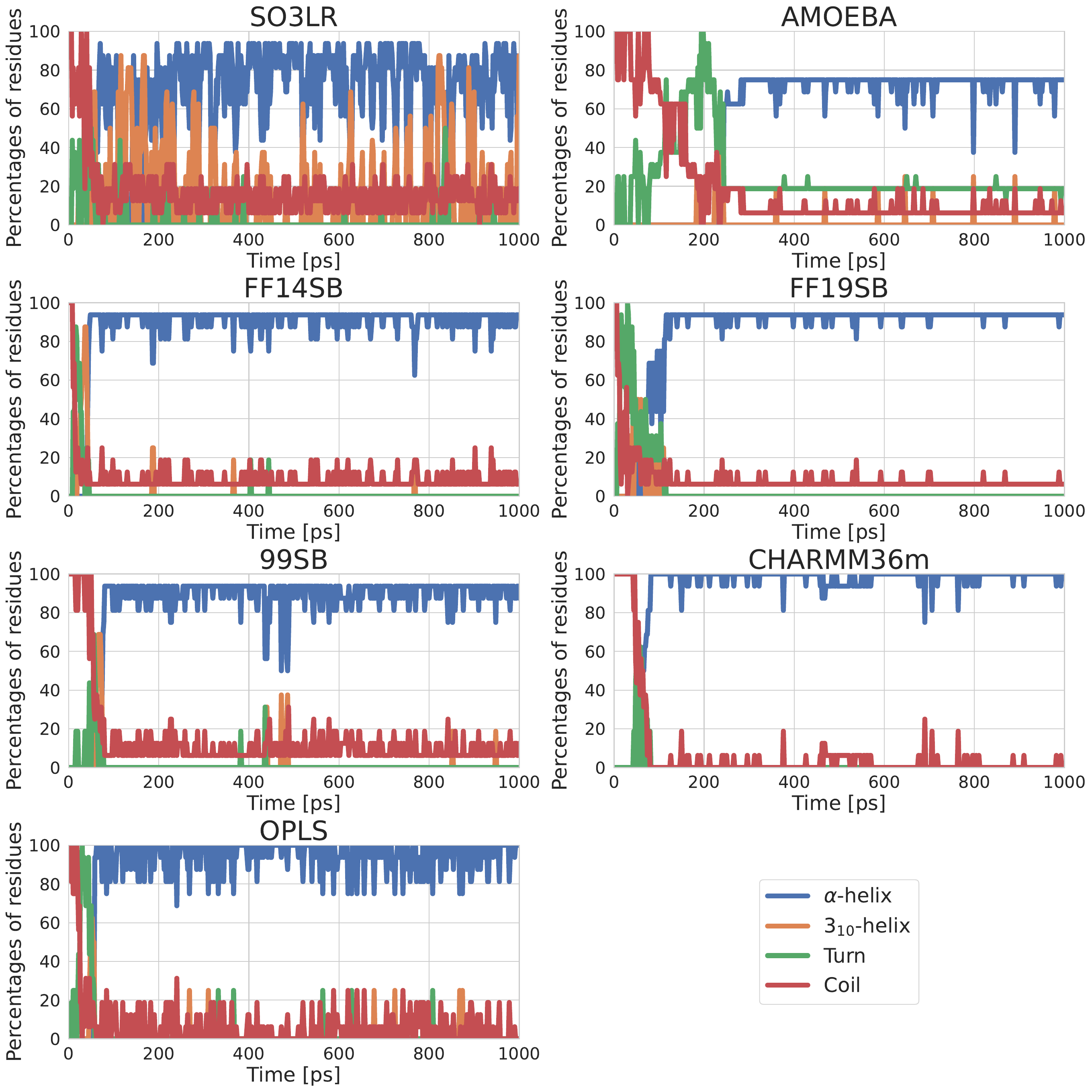}
\caption{AceAla15NMe peptide folding path of different FF from complete extended peptide to helical folding during 1 ns of trajectory. Secondary structure was computed with STRIDE \cite{Frishman1995}.}
\label{sfig:alafold}
\end{figure*}

\begin{figure*}[b]
\centering
\includegraphics[width=\linewidth]{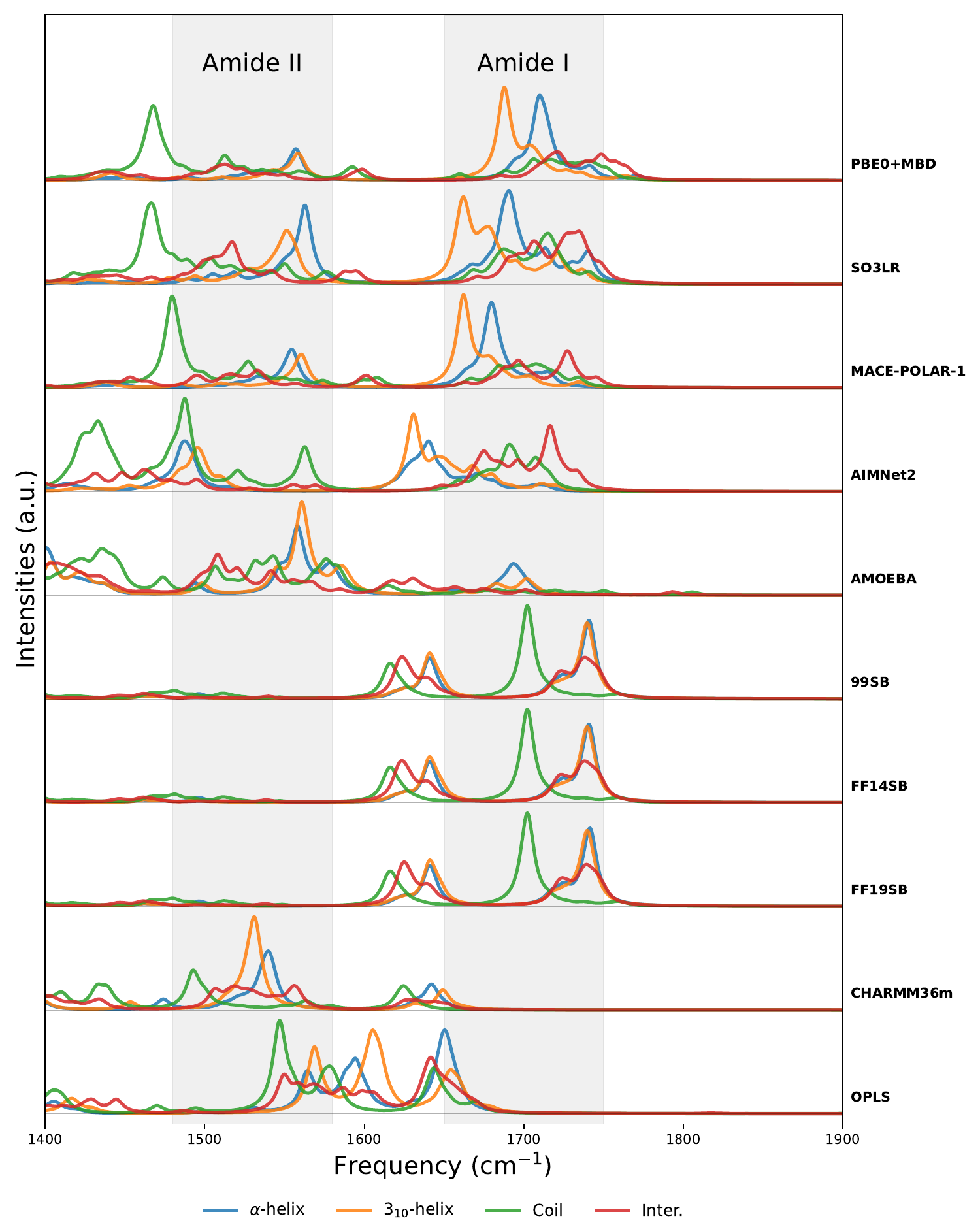}
\caption{AceAla\textsubscript{15}NMe peptide infrared spectrum through harmonic approximation of 4 different conformers, canonical $\alpha$-helix (blue), canonical 3\textsubscript{10}-helix (orange), fully extended (Coilm green). and intermediate folding state (Inter., red). Reference DFT calculation at PBE0+MBD level. }
\label{sfig:alafold}
\end{figure*}

\begin{figure*}[b]
\centering
\includegraphics[width=\linewidth]{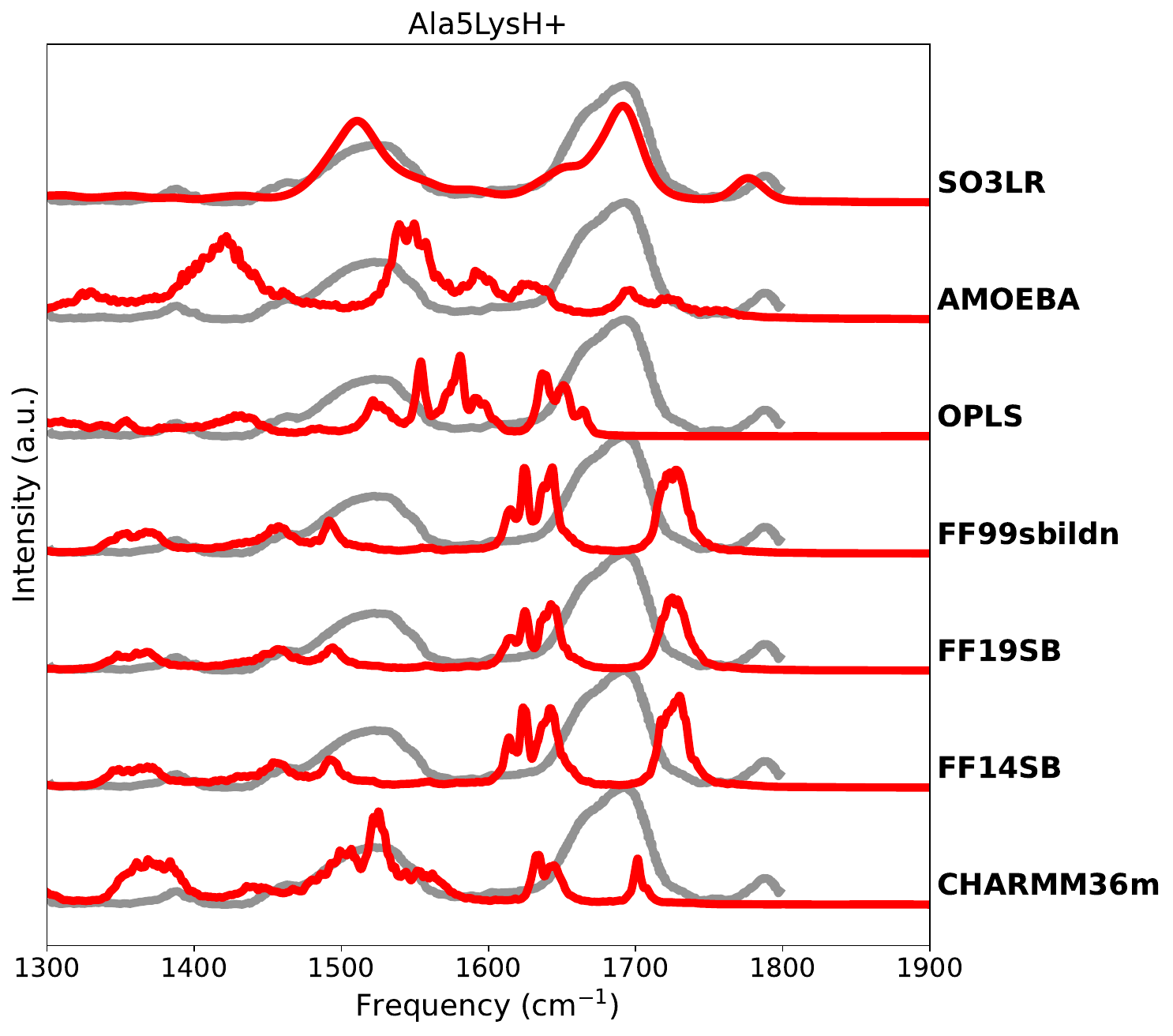}
\caption{Computed infrared spectrum of AceAla\textsubscript{5}LysH\textsuperscript{+} obtained from dipole autocorrelation analysis of 20 × 100 independent NVE molecular‑dynamics simulations, compared with experimental data from Ref.~\cite{Rossi2010}.}
\label{sfig:alafold}
\end{figure*}

\begin{figure*}[b]
\centering
\includegraphics[width=\linewidth]{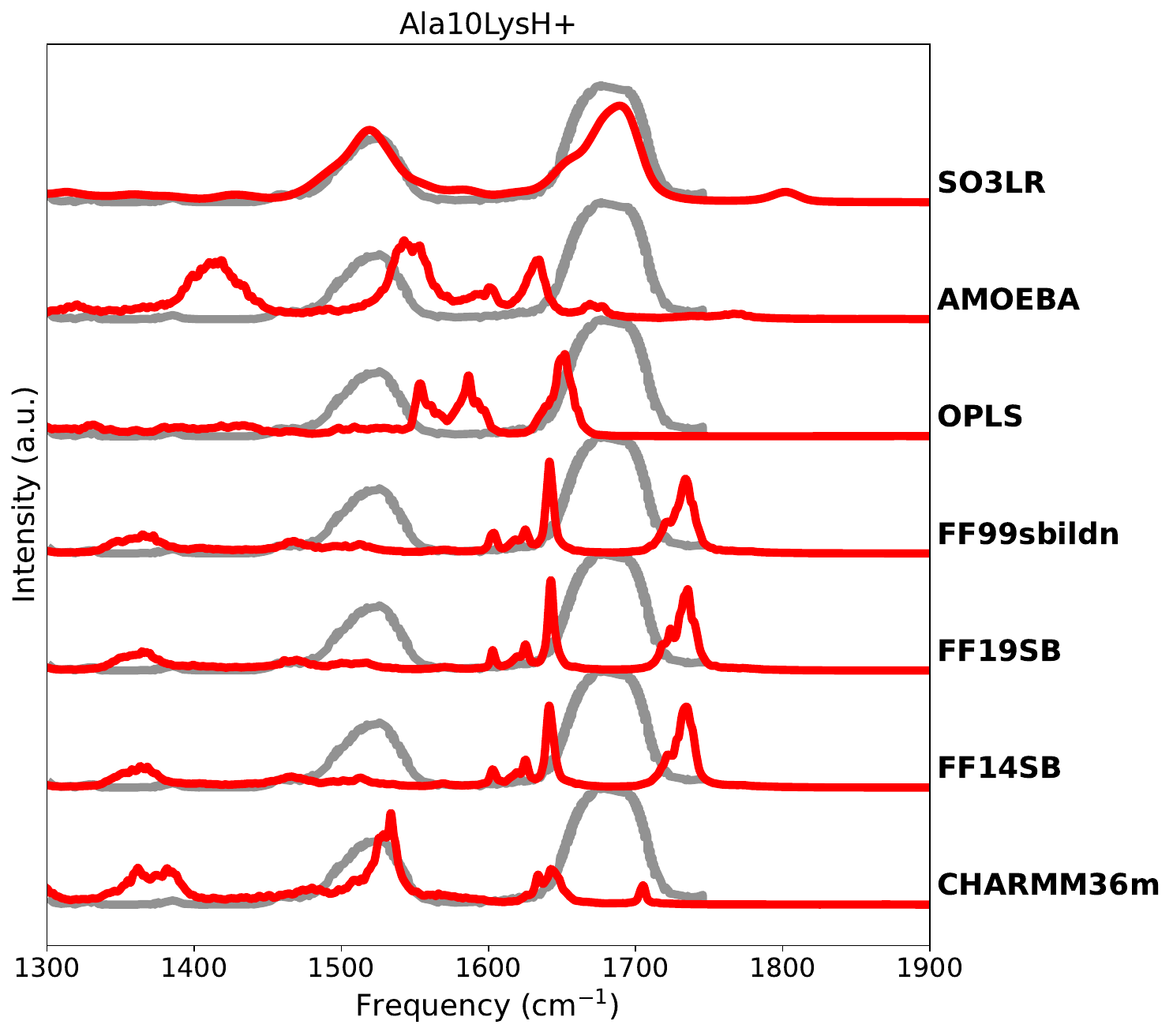}
\caption{Computed infrared spectrum of AceAla\textsubscript{10}LysH\textsuperscript{+} obtained from dipole autocorrelation analysis of 20 × 100 independent NVE molecular‑dynamics simulations, compared with experimental data from Ref.~\cite{Rossi2010}.}
\label{sfig:alafold}
\end{figure*}

\begin{figure*}[b]
\centering
\includegraphics[width=\linewidth]{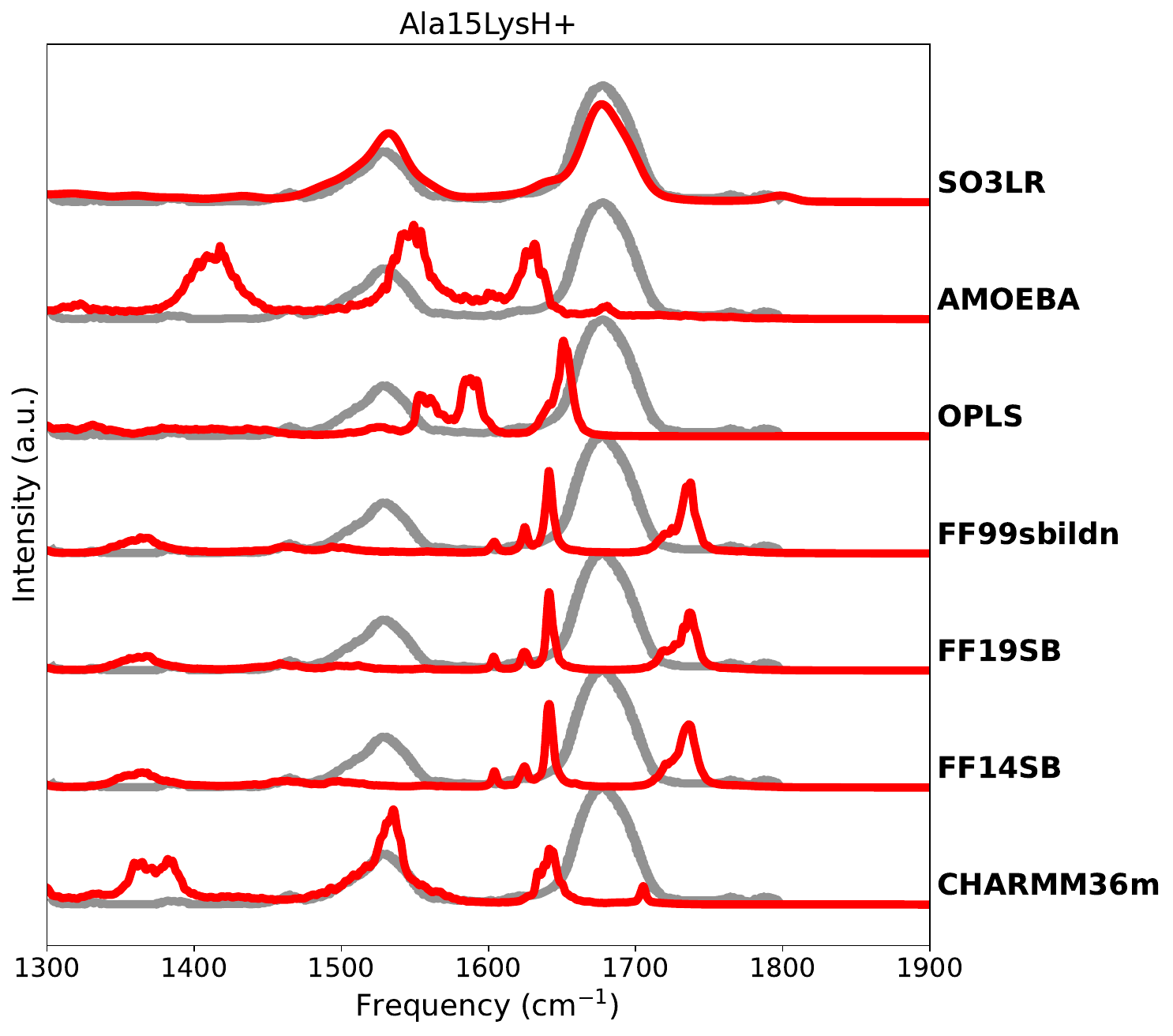}
\caption{Computed infrared spectrum of AceAla\textsubscript{15}LysH\textsuperscript{+} obtained from dipole autocorrelation analysis of 20 × 100 independent NVE molecular‑dynamics simulations, compared with experimental data from Ref.~\cite{Rossi2010}.}
\label{sfig:alafold}
\end{figure*}

\begin{figure*}[b]
\centering
\includegraphics[width=\linewidth]{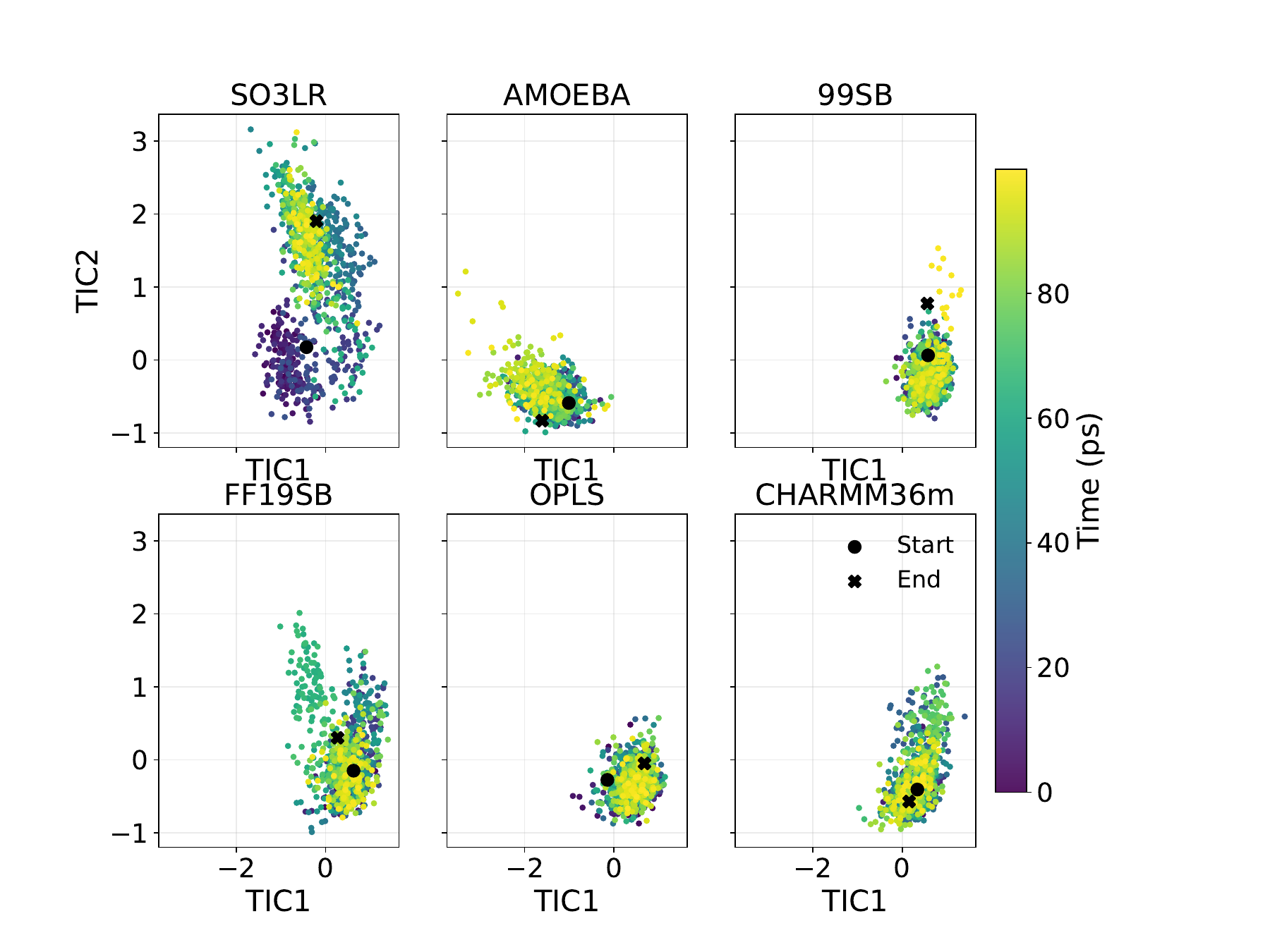}
\caption{Time-resolved trajectories in the common TICA \cite{Perez-Hernandez2013} space for Ala$_5$LysH$^{+}$ simulated with different force fields (1 replica, 100ps). A single TICA model was fit to pooled backbone $\phi$ and $\psi$ dihedral features encoded as $\sin$ and $\cos$ terms. Points are colored by simulation time. Circle and cross markers indicate the initial and final conformations, respectively.}
\label{sfig:alafold}
\end{figure*}

\begin{figure*}[b]
\centering
\includegraphics[width=\linewidth]{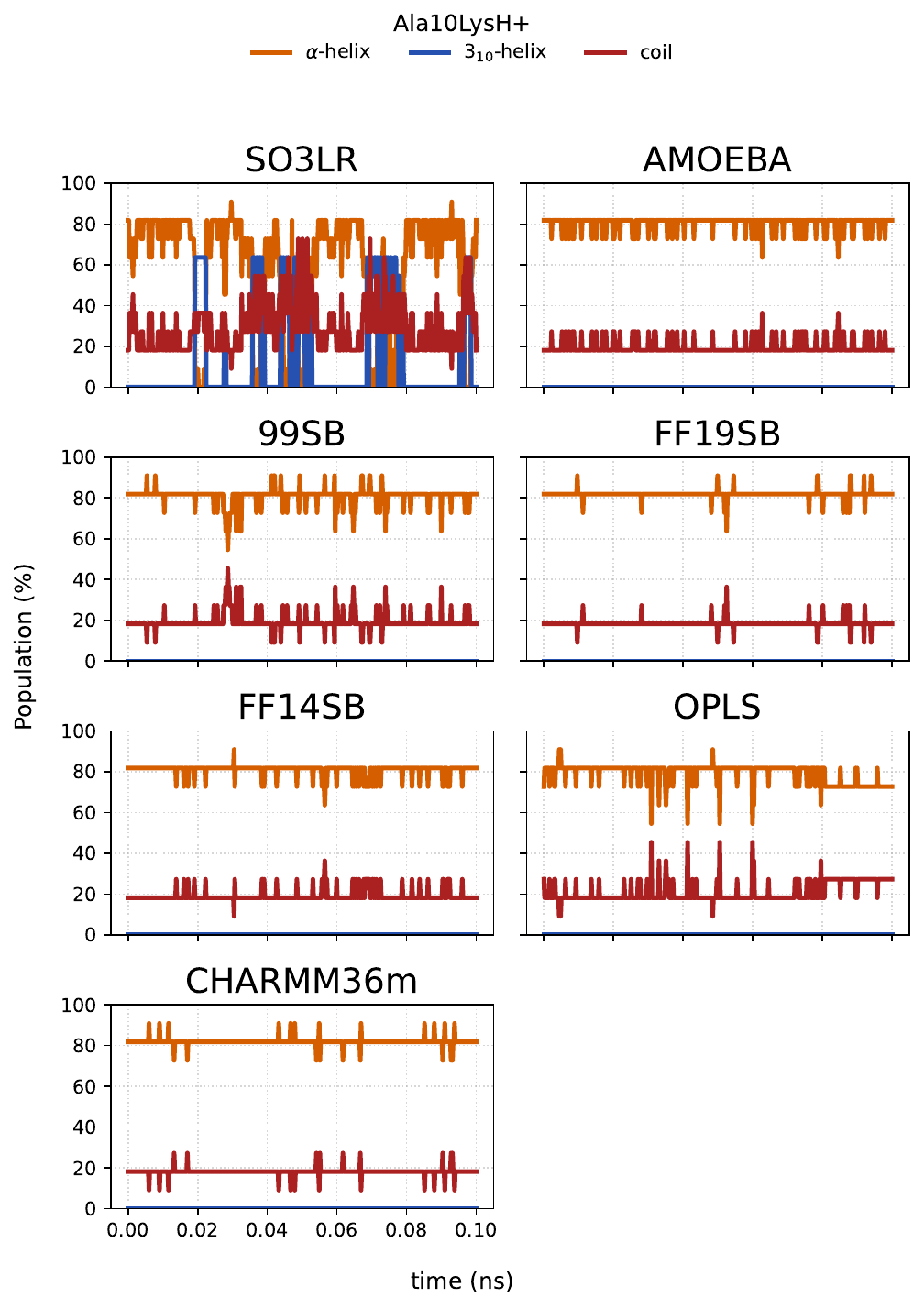}
\caption{Ala$_{10}$LysH$^{+}$ peptide secondary structure over 100ps MD. Secondary structure was computed with STRIDE \cite{Frishman1995}.}
\label{sfig:ala10lysfold}
\end{figure*}

\begin{figure*}[b]
\centering
\includegraphics[width=\linewidth]{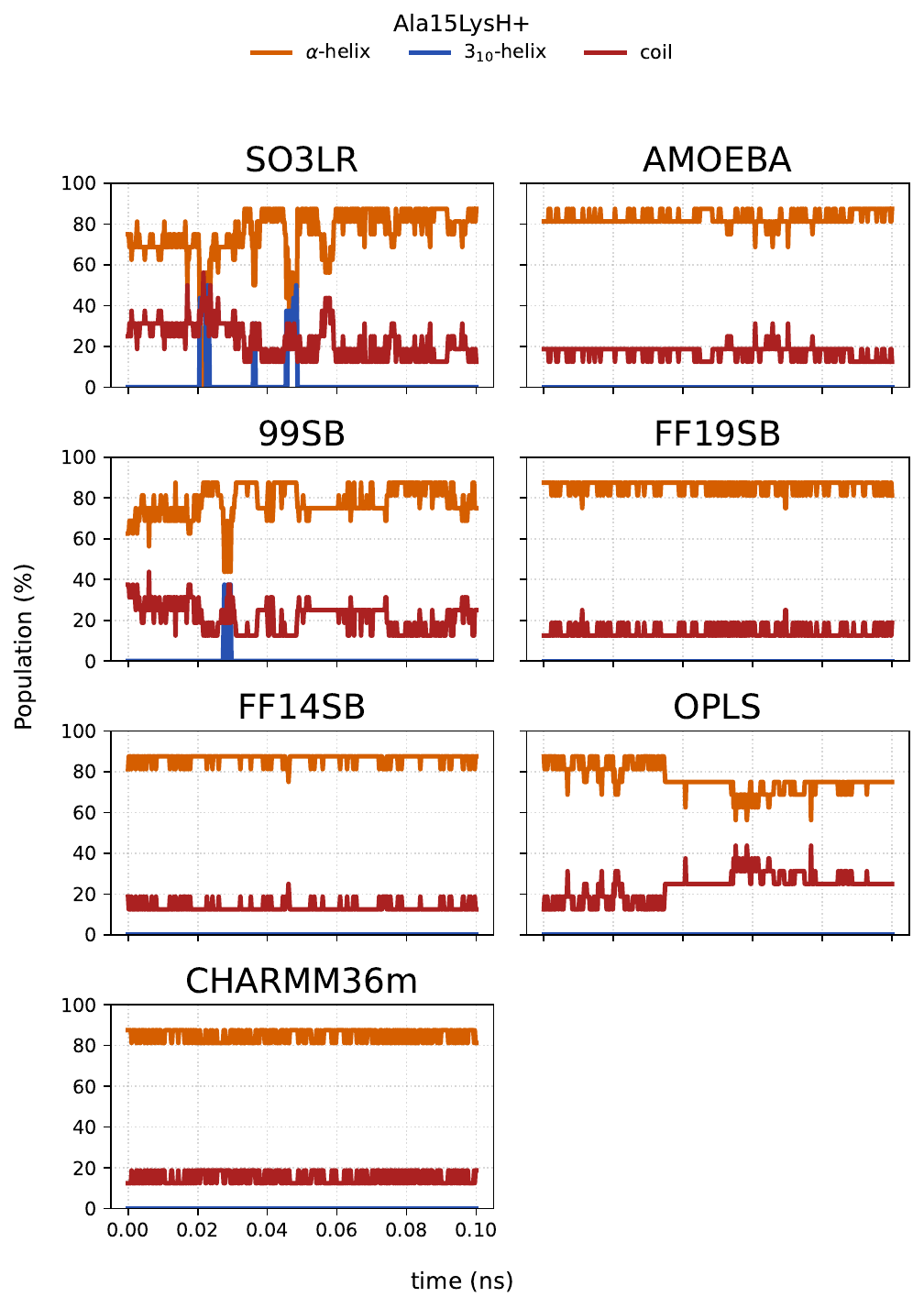}
\caption{Ala$_{15}$LysH$^{+}$ peptide secondary structure over 100ps MD. Secondary structure was computed with STRIDE \cite{Frishman1995}.}
\label{sfig:ala10lysfold}
\end{figure*}

\begin{figure}[b]
\centering
\includegraphics[width=\linewidth]{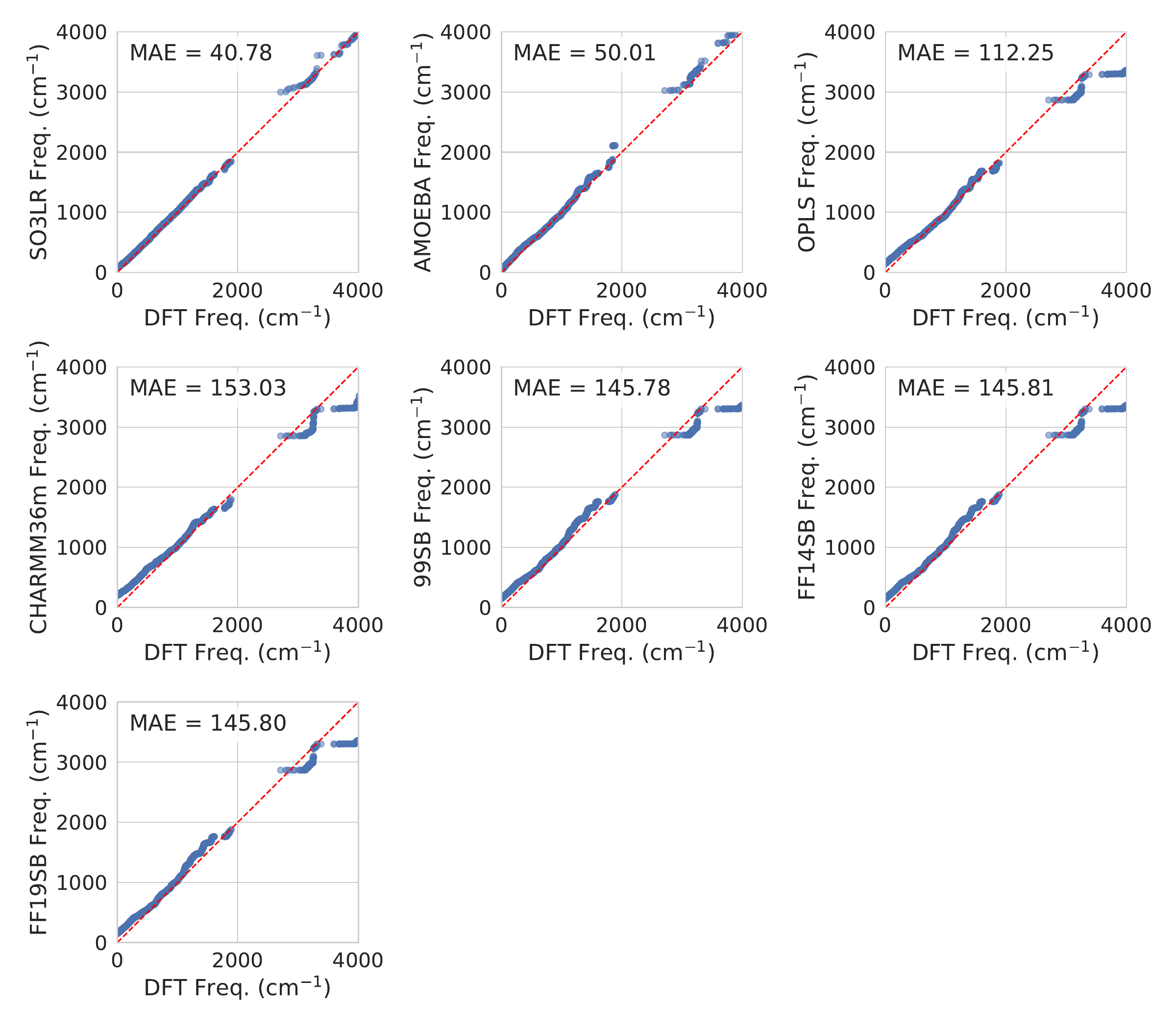}
\caption{Vibrational frequency analysis of p53 monomer. SO3LR and 6 different FFs frequencies were compared to those from DFT at PBE+MBD level. Mean absolute error (MAE) from 1 to 1 fitting is shown.}
\label{sfig:p53freq}
\end{figure}

\begin{figure}[b]
\centering
\includegraphics[width=\linewidth]{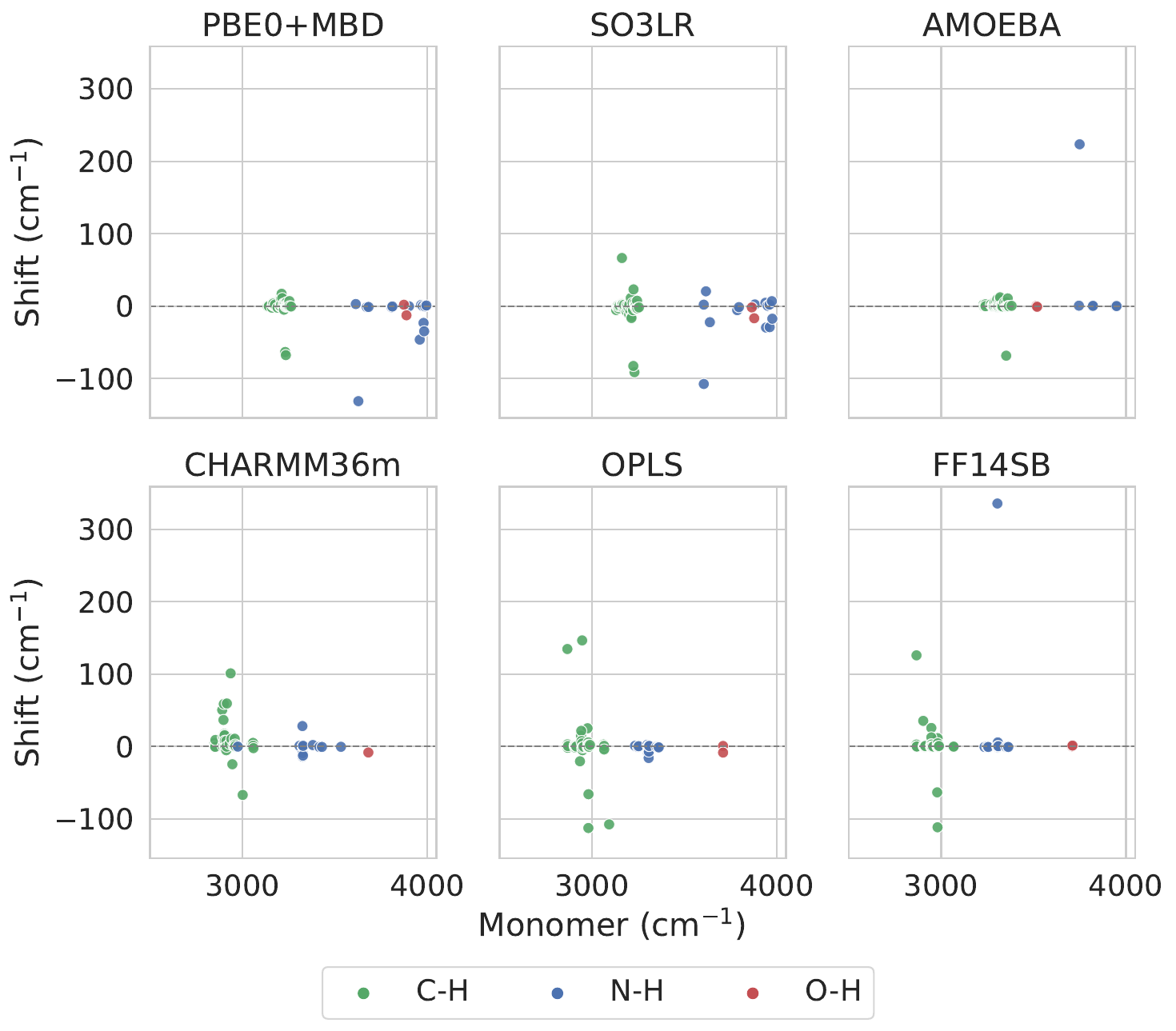}
\caption{High-frequency vibrational analysis of the $\beta$-sheet of p53 dimer (residues 227-333). Vibrational shifts (cm\textsuperscript{-1}) of high-frequency modes in dimer versus monomer. The x-axis is the monomer frequencies (cm\textsuperscript{-1}). Each point corresponds to the most similar mode matched by the atom index. Vibrational modes are color-coded as follows: green for C–H stretching, blue for N–H stretching and red for O–H stretching.}
\label{sfig:betascatter}
\end{figure}

\begin{figure}[b]
\centering
\includegraphics[width=\linewidth]{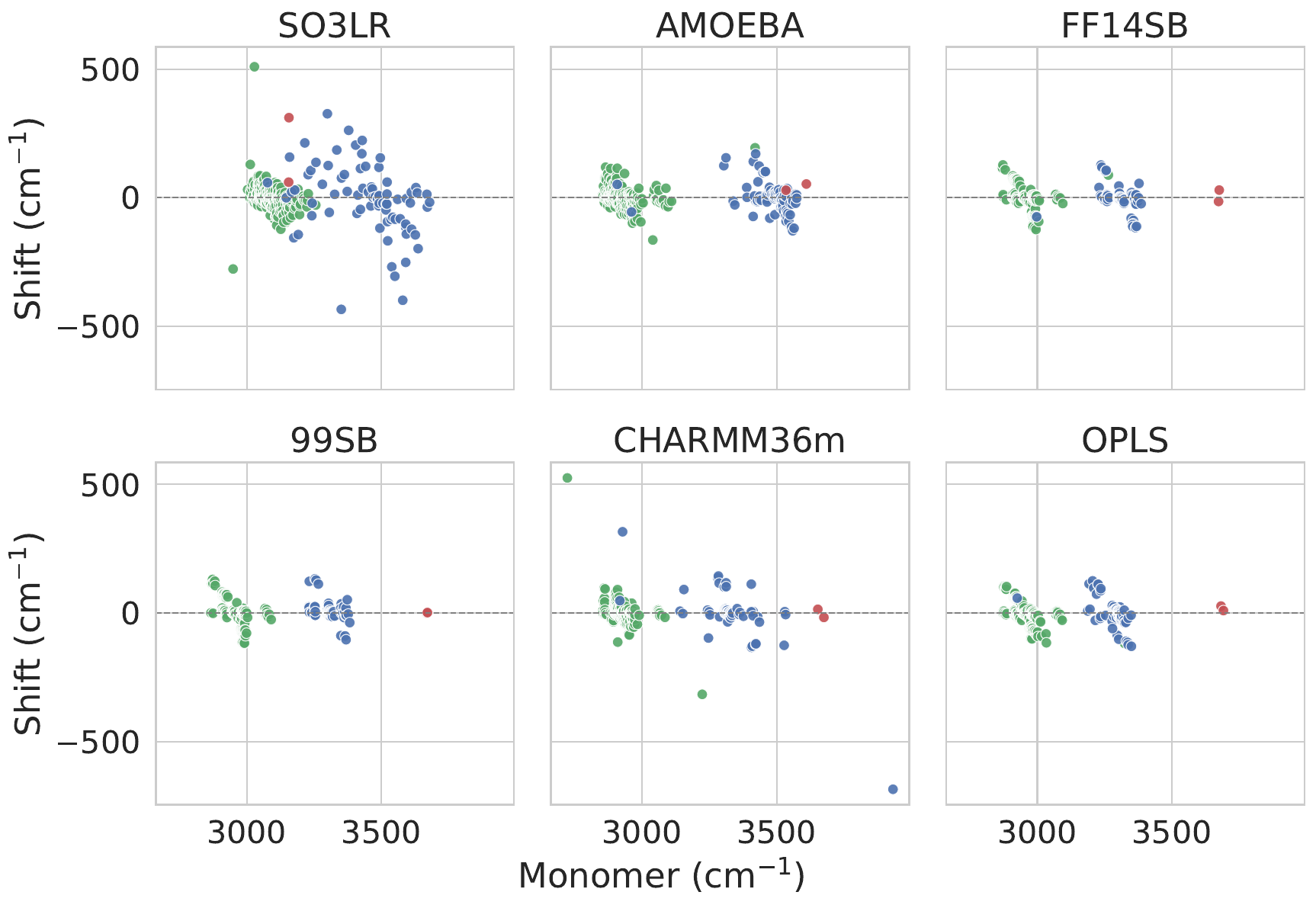}
\caption{High-frequency vibrational analysis of p53 tetramer and monomer in water. Vibrational shifts (cm\textsuperscript{-1}) of high-frequency modes in dimer versus monomer. The x-axis is the monomer frequencies (cm\textsuperscript{-1}). Each point corresponds to the most similar mode matched by the atom index. Vibrational modes are color-coded as follows: green for C–H stretching, blue for N–H stretching and red for O–H stretching.}
\label{sfig:p53tetramer}
\end{figure}

\begin{figure}[b]
\centering
\includegraphics[width=\linewidth]{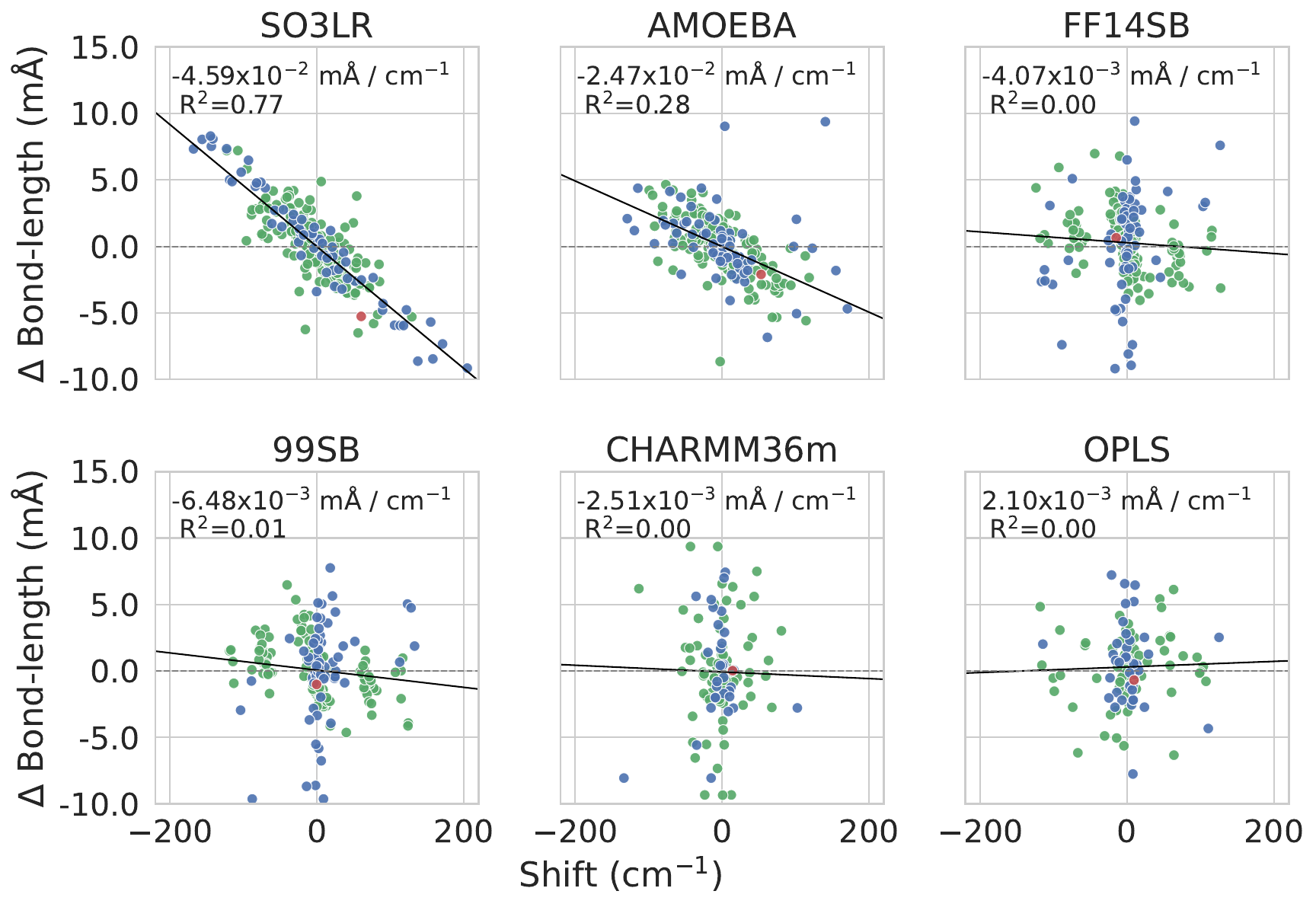}
\caption{Vibrational shifts plotted against bond‑length differences between solvated tetramer and monomer configurations, including the linear correlation $R^2$ and the m\AA/cm\textsuperscript{-1} slope. Vibrational modes are color-coded as follows: green for C–H stretching, blue for N–H stretching and red for O–H stretching.}
\label{sfig:p53correlation}
\end{figure}

\clearpage
\section{Amino Acids under Solvent Effects}\label{Aminoacids}
\setcounter{figure}{0}

Tyrosine, Leucine, and Arginine, each capped with Ace and NMe groups, were studied both in vacuum and in a water shell extending 3\AA~ from the molecular atoms. The amino acids in vacuum were optimized using the PBE0+MBD functional with 'tight' settings, following the same approach described in the small molecule section. Subsequently, a larger shell of water molecules (8\AA) was added to the optimized conformations. An energy minimization was performed using xtb \cite{Bannwarth2021} with the amino acid atoms fixed, allowing the surrounding water molecules to relax. After this step, a 3\AA~ water shell was extracted. A second minimization cycle was then carried out, again constraining the amino acid atoms.

To ensure a fair comparison, the same conformers, both in vacuum and in water, were used across all methods. Normal mode analysis was performed as described in the Methods section.

\subsection{Tyrosine}

\begin{figure}[H]
\centering
\includegraphics[width=\linewidth]{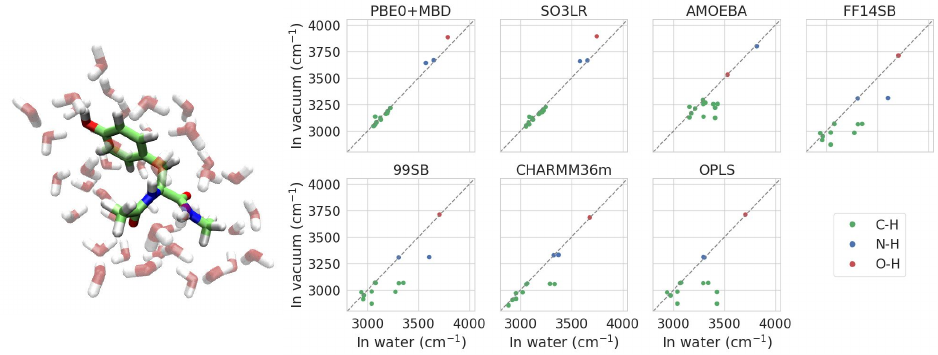}
\caption{Vibrational frequency shifts of high-frequency modes in water (x-axis) compared to vacuum (y-axis) for the Ace-Tyr-NMe amino acid, computed using PBE0+MBD, SO3LR, and the most commonly used empirical molecular force fields.}
\label{sfig:tyr}
\end{figure}

\subsection{Leucine}

\begin{figure}[H]
\centering
\includegraphics[width=\linewidth]{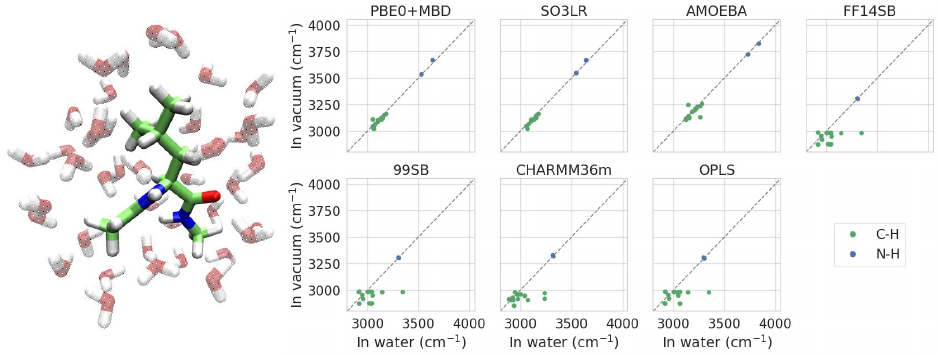}
\caption{Vibrational frequency shifts of high-frequency modes in water (x-axis) compared to vacuum (y-axis) for the Ace-Leu-NMe amino acid, computed using PBE0+MBD, SO3LR, and the most commonly used empirical molecular force fields.}
\label{sfig:leu}
\end{figure}

\subsection{Arginine}

\begin{figure}[H]
\centering
\includegraphics[width=\linewidth]{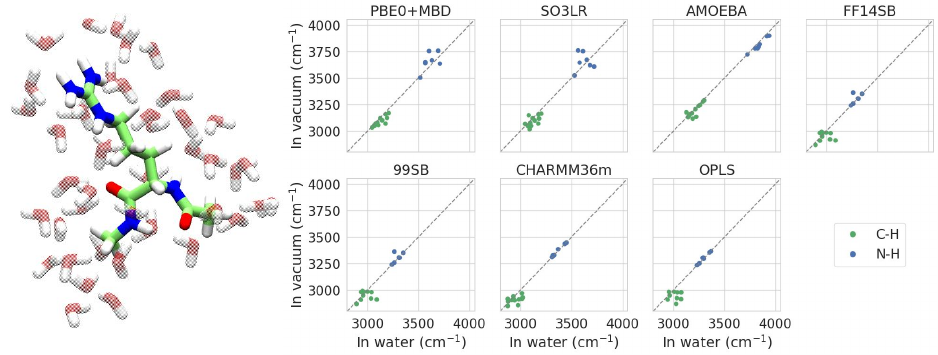}
\caption{Vibrational frequency shifts of high-frequency modes in water (x-axis) compared to vacuum (y-axis) for the Ace-Arg-NMe amino acid, computed using PBE0+MBD, SO3LR, and the most commonly used empirical molecular force fields.}
\label{sfig:arg}
\end{figure}
\end{appendices}
\bibliography{references.bib}